\documentclass[fleqn,usenatbib]{mnras}

\usepackage{newtxtext,newtxmath}

\usepackage[T1]{fontenc}
\usepackage{ae,aecompl}

\usepackage{graphicx}	
\usepackage{amsmath}	
\usepackage{amssymb}	

\usepackage{color}
\usepackage{mathrsfs}
\usepackage{wasysym}

\providecommand{\e}[1]{\ensuremath{\times 10^{#1}}}

\newcommand{\md}{\mathrm{d}}
\renewcommand{\eqref}[1]{equation~(\ref{#1})}
\newcommand{\tabref}[1]{Table~\ref{#1}}
\newcommand{\appref}[1]{Appendix~\ref{#1}}
\newcommand{\xuabbr}{Xu+16 }
\newcommand{\xuabbrns}{Xu+16}
\newcommand{\xucf}[1]{Cf. Fig.~#1 of Xu+16.}
\newcommand{\xucftab}[1]{Cf. Table~#1 of Xu+16.}
\newcommand*{\bfrac}[2]{\genfrac{}{}{0pt}{}{#1}{#2}}

\title[Reducing biases on $H_0$]{Reducing biases on $H_0$ measurements using strong lensing and galaxy dynamics: results from the EAGLE simulation}
\author[Tagore et al.]
{Amitpal S. Tagore$^{1}$, David J. Barnes$^{1}$, Neal Jackson$^{1}$, Scott T. Kay$^{1}$, \\
\mbox{}\\
{\rm \LARGE Matthieu Schaller$^{2}$, Joop Schaye$^{3}$, Tom Theuns$^{2}$}\\
\mbox{}\\
$^{1}$Jodrell Bank Centre for Astrophysics, School of Physics \& Astronomy, 
University of Manchester, Turing Building, Oxford Road, \\
Manchester M13 9PL, UK\\
$^{2}$Institute for Computational Cosmology, Department of Physics, Durham University, South Road, Durham, DH1 3LE, UK\\
$^{3}$Leiden Observatory, Leiden University, PO Box 9513, 2300 RA, Leiden, the Netherlands \\
}

\date{Accepted XXX. Received YYY; in original form ZZZ}

\pubyear{2017}

\begin{document}
\label{firstpage}
\pagerange{\pageref{firstpage}--\pageref{lastpage}}
\maketitle

\begin{abstract}
Cosmological parameter constraints from observations of time-delay lenses are becoming increasingly precise.
However, there may be significant bias and scatter in these measurements due to, among other things, the so-called mass-sheet degeneracy.
To estimate these uncertainties, we analyze strong lenses from the largest \textsc{eagle} hydrodynamical simulation.
We apply a mass-sheet transformation to the radial density profiles of lenses, and by selecting lenses near isothermality, we find that the bias on $H_0$ can be reduced to 5\% with an intrinsic scatter of 10\%, confirming previous results performed on a different simulation data set.
We further investigate whether combining lensing observables with kinematic constraints helps to minimize this bias.
We do not detect any significant dependence of the bias on lens model parameters or observational properties of the galaxy, but depending on the source--lens configuration, a bias may still exist.
Cross lenses provide an accurate estimate of the Hubble constant, while fold (double) lenses tend to be biased low (high).
With kinematic constraints, double lenses show bias and intrinsic scatter of 6\% and 10\%, respectively, while quad lenses show bias and intrinsic scatter of 0.5\% and 10\%, respectively.
For lenses with a reduced $\chi^2>1$, a power-law dependence of the $\chi^2$ on the lens environment (number of nearby galaxies) is seen.
Lastly, we model, in greater detail, the cases of two double lenses that are significantly biased.
We are able to remove the bias, suggesting that the remaining biases could also be reduced by carefully taking into account additional sources of systematic uncertainty.
\end{abstract}


\section{Introduction}
\label{sec:intro}

Strong gravitational lensing has long played an important role in astronomy.
In strongly lensed systems, the magnification of the lensed source can allow for detailed studies of the source and the mass distribution of the lens.
It can also place constraints on cosmological parameters that are independent from those of other methods, such as lensing of the cosmic microwave background and supernovae distance measurements \citep[see][for a comparison of methods]{nealcosm,neal2}.
By continuously monitoring the lensed images of a time-variable source, such as a quasar, the delays in arrival of photons at the image locations can be measured, which in turn are relatable to cosmology \citep{refsdal}.
The time of arrival of photons at a position in the lens plane, $\vec{x}$, is given by
\begin{equation}
t(\vec{x}) = \frac{1+z_d}{c}\frac{D_d D_s}{D_{ds}}\bigg(\frac{1}{2}|\vec{x}-\vec{u}|^2-\phi(\vec{x})\bigg),
\end{equation}
where $z_d$ is the redshift of the lens; $c$ is the speed of light; $D_d$, $D_s$, and $D_{ds}$ are, respectively, the angular diameter distances from the observer to lens, observer to source, and lens to source; $\vec{u}$ is the position of the unlensed source; and $\phi(\vec{x})$ is the dimensionless lens potential.
Because of the dependence of the time delay on ratios of cosmological distances, these measurements are particularly sensitive to the Hubble constant, $H_0$.

In the late 1990s the first measurements of time delays and inferences of the Hubble constant were made \citep[see, e.g.][]{lens1,lens2,lens3,lens4,firstlenses}.
The density profiles of these early lenses were not always strongly constrained, but the addition of kinematic information \citep{lens5} and emission from the quasar host galaxy \citep{lens6} helped to remove degeneracies present in the lens modelling.
More recently, advanced techniques and more detailed modelling have increased the reliability of strong lensing measurements.
The $H_0$ Lenses in COSMOGRAIL Wellspring (H0LiCOW) program \citep{holicow:i} are now using time delay measurements from the COSmological MOnitoring of GRAvItational Lenses \citep[COSMOGRAIL;][]{cosmograil} to make precision measurements of the Hubble constant.
The H0LiCOW program addresses several significant systematics in strong lens modelling by identifying galaxies in the group of the lens or along the line of sight \citep{holicow:ii}, quantifying effects of mass along the line-of-sight \citep{holicow:iii}, minimizing confirmation bias through blind lens modelling, and utilizing single and multi-component lens models \citep{holicow:iv}.
Combining measurements from three time delay lenses, \citet{holicow:v} find, for the case of a $\Lambda\textnormal{CDM}$ cosmology, a Hubble constant of $H_0=71.9^{+2.4}_{-3.0}\mathrm{km}\,\mathrm{s}^{-1}\mathrm{Mpc}^{-1}$.
This result is independent of any other method and is not in significant tension with other probes.
For example, the measurement agrees with the \textit{Planck} $2015$ results at the 1--2$\sigma$ level.
The authors also go on to explore other cosmological models, constraining the curvature parameter $k$ and the dark energy equation of state $w$ as well.

Given the increasing quality of data and modelling techniques, tests of the methods using numerical simulations are critical for understanding possible sources of bias.
A well-known source of uncertainty is the so-called mass-sheet degeneracy \citep[MSD;][]{MSD1,MSD2}.
Under the MSD, a given convergence profile $\kappa(\vec{x})=\Sigma(\vec{x})/\Sigma_\textnormal{cr}$\footnote{$\Sigma(\vec{x})$ is the projected surface mass density, and $\Sigma_\textnormal{cr}$ is the critical surface density for lensing.} can be transformed into another convergence given by
\begin{equation}
\kappa_\lambda(\vec{x}) = \lambda \kappa(\vec{x}) + (1-\lambda),
\label{eq:eq1}
\end{equation}
where $\lambda$ is a constant.
This transformation leads to a simultaneous, but unobservable, transformation of the unlensed source properties and no change in the positions and fluxes of lensed images.
However, the product of the Hubble constant and time delays is affected such that $H_0\Delta t\rightarrow\lambda H_0\Delta t$.
Thus, the inferred value of the Hubble constant will be biased by the factor $\lambda$. 
In practice it is typical for lenses to be modelled using power-law density profiles, and because power-laws do not strictly map to power-laws under the mass-sheet transformation (MST), they mathematically break this degeneracy.
However, in doing so, they artificially pick out a particular transformation among many possible solutions, leading to a direct bias on the value of $H_0$ inferred from such a model.
Additionally, independent constraints on the mass profile, such as those from velocity dispersion measurements, can help break the degeneracy and minimize this bias, but may also introduce additional systematic uncertainties.

Until recently, cosmological {\it N}-body simulations have not been able to realistically model galaxy-scale lenses.
Dark matter only simulations, such as Millenium-XXL \citep{millxxl}, simulate large cosmological volumes, but the resolution is limited by the mass of the dark matter particles and the gravitational softening length.
For the Millenium-XXL project, these correspond to particles with masses of $8.5\e{9} \rm{M}_{\astrosun}$ and a softening length of 13.7 kpc, which are not small enough to resolve the structure of galaxies.
Moreover, dark matter only simulations do not take into account the effects of baryons, which are a key component to analyzing strong lenses since the Einstein radius is typically within the region where baryons and dark matter are both present in significant amounts.
 
State of the art simulations can now model both the baryons and dark matter in galaxies, reproducing a wide range of their observed properties.
This increase in resolution and astrophysical modelling comes at the cost of a smaller simulation box.
Whereas the Millenium-XXL simulation was 4.1 Gpc on each side, baryon and dark matter simulations are typically done in $\sim 100$ Mpc boxes.
Recent efforts include the Illustris project \citep{illustris}, the \textsc{eagle} project \citep{eagle1,eagle2}, the \textsc{mufasa} project \citep{mufasa}, and the \textsc{romulus} simulations \citep{romulus}.
Although there are many similarities between the simulations, there are several key differences.
The hydrodynamic scheme to simulate the fluid elements varies, with \textsc{eagle} and \textsc{romulus} using smooth-particle hydrodynamics (SPH), \textsc{mufasa} using meshless, finite-mass hydrodynamics and Illustis using a Voronoi tessellation adaptive mesh scheme.
The codes used to solve for the gravitational interactions between the particles or fluid elements also differ between the simulations.
\textsc{romulus} specifically aims to capture the detailed formation and evolution of super-massive black holes with better subgrid models.
All of the simulations are calibrated to reproduce some particular property of present-day galaxies.
Illustris, \textsc{eagle}, and \textsc{mufasa} were calibrated to reproduce the observed low-redshift galaxy stellar mass function, while \textsc{romulus} calibrated on the observed stelar mass--halo mass relationship.
However, the \textsc{eagle} project is the only simulation which is specifically calibrated to reproduce the observed low-redshift galaxy mass--size relationship.
Since the radial profile of a galaxy and the concentration of matter within its central region play an important role in determining its lens properties, \textsc{eagle} galaxies are especially well-suited to investigate strong lenses.

Recently, \citet{xubias}, hereafter \xuabbrns, have examined the average radial profile of galaxies in the Illustris simulation.
The authors extract the convergence at two different radii (representing the positions of two lensed images) and, assuming power-law density profiles for the lens, calculate the average density slope between the two images.
The convergence at the midpoint is also calculated, and then an MST is applied to this density slope so that the three points (the lensed image locations and the midpoint) lie on a line in log--log space.
We note that although the mass-sheet degeneracy can be thought of as being due to a uniform sheet of mass at the redshift of the lens, there are many manifestations of the degeneracy.
In particular, \xuabbr focus on local deviations of the mass density from a power-law near lensed images.
In observations, these local deviations would lead to an inferred power-law slope different from the average slope between the two lensed images, introducing a multiplicative bias on $H_0$.

In practice, the convergence at the three aforementioned radial positions is not directly observed; positions, fluxes, and time delays are the primary observables.
Still, with rudimentary lensing information alone, there exists a strong degeneracy between the mass inside the Einstein radius, a robustly determined quantity for fixed $H_0$, and $H_0$ itself.
Here, we use mock observations of lens galaxies in the \textsc{eagle} simulation to assess the ability to recover $H_0$ given positions of lensed images, time delays, and velocity dispersion information.
We also investigate how adding additional information from the extended light distribution of the quasar host galaxy and from the lens environment can further help to break degeneracies and to minimize bias.

The remainder of this paper is outlined as follows.
Section~\ref{sec:eagle} provides a brief description of the \textsc{eagle} simulation.
Section~\ref{sec:comparison} applies an MST to the radial profiles of \textsc{eagle} lenses to assess possible biases on $H_0$ measurements.
The results are then compared to those of \xuabbr.
In section~\ref{sec:modelling}, we use lensing observables and kinematic constraints to constrain $H_0$.
We address the possible effects of lens environment and constraints from the host galaxy in section~\ref{sec:adv}.
Finally, we summarize our findings in section~\ref{sec:conclusion}.

\section{The \textsc{eagle} project}
\label{sec:eagle}
\textsc{eagle} is a project of the Virgo Consortium.
It is a suite of cosmological hydrodynamical simulations of periodic cubic volumes designed to study galaxy formation and evolution.
The \textsc{eagle} code is a modified version of \textsc{p-gadget-3}, which is an updated version of \textsc{p-gadget-2} \citep{gadget}.
We focus on the reference model in the volume with a comoving side length of $100\,\mathrm{Mpc}$, as this contains the largest sample of possible lenses.
This volume assumes a $\Lambda\mathrm{CDM}$ cosmology with parameters taken from the \textit{Planck} $2013$ results \citep{planck2014i}: $\Omega_{\mathrm{b}}=0.0483$, $\Omega_{\mathrm{M}}=0.307$, $\Omega_{\Lambda}=0.693$, $h=0.6777$, $\sigma_{8}=0.8288$, $n_{\rm{s}}=0.9611$.
Below, we briefly describe the subgrid physics of the \textsc{eagle} model.

Radiative cooling and photoheating is implemented following \citet{wiersma}, assuming a \citet{haardt} optically thin X-ray/UV background.
Star formation is implemented in a stochastical manner following \citet{sv08}, which by construction reproduces the observed Kennicutt-Schmidt law.
Stars form at a pressure-dependent rate above a metallicity-dependent density threshold.
Each star particle is assumed to be a simple stellar population with a \citet{chabrier} initial mass function in the range $0.1-100\,\mathrm{M}_{\astrosun}$.

Stellar evolution is modelled following \citet{wiersma}, where the metallicity-dependent release of $11$ chemical elements from AGB stars and Type Ia and Type II supernovae is tracked.
Stellar feedback is implemented by stochastic heating particles by a fixed temperature increment \citep{dvs2012}.
The seeding, growth and feedback from super massive black holes (BHs) is based on \citet{sdh2005} with modifications from \citet{bs2009} and \citet{rosas}.
Feedback from BHs is implemented as a single mode.

A critical aspect of state-of-the-art galaxy formation models is the calibration of the subgrid physics.
The \textsc{eagle} project calibrated the free parameters associated with stellar feedback to reproduce the observed low-redshift galaxy stellar mass function and the observed low-redshift galaxy mass-size relation in the stellar mass range $10^{9}-10^{11}\,\mathrm{M}_{\astrosun}$.
After this calibration the simulation reproduces the observed evolution of both the galaxy mass function and galaxy sizes \citep{Furlong15,Furlong17}.
The \textsc{eagle} galaxies are ideally suited for a strong lens study as their stellar mass and their extent are a good match to observational constraints.
Therefore, they should provide a more realistic lens population compared to previous simulations.

\section{Comparison to previous work}
\label{sec:comparison}

\xuabbr have used the Illustris simulation to show that there can be a very strong bias and large scatter in measurements of $H_0$ from strong lensing.
It is not immediately clear, however, if the results are dependent on the choice of simulation.
Here, we perform a similar analysis using the \textsc{eagle} simulation, focusing on minimizing the differences between the two analyses.

\subsection{Extracting lens properties}
\label{sec:lens_selection}

To best match the lens criteria of \xuabbr and to ensure that only well-resolved, realistic lens candidates are extracted from the redshift snapshots listed in \tabref{tab:galstat}, several selection cuts are applied.
For details of the calculations and methods described in this section, see \appref{app:pdp1}.
First, a lower-limit, friends-of-friends\footnote{The friends-of-friends method identifies halos by including in the halo all dark matter particles within a linking length of 0.2 times the mean particle separation. Baryonic particles are assigned to the same halo, if any, to which their nearest dark matter particle is assigned.} mass cut of $10^{11} \rm{M}_{\astrosun}$ is applied. 
Then, three projections along the coordinate axes give three potential lens candidates for each galaxy in the simulation.
A galaxy is accepted as a lens if its circularized Einstein radius\footnote{The angle within which the mean convergence is unity.} is more than twice the gravitational softening length ($2\times 700$ pc).
A final selection cut is made, requiring the one-dimensional velocity dispersion ($\sigma_\textnormal{SIS}$) to be greater than 160~$\mathrm{km}\,\mathrm{s}^{-1}$.\footnote{We assume a circular, isothermal lens, so that the Einstein radius is given by $4\pi(\sigma_\textnormal{SIS}/c)^2 D_d/D_{ds}$, where $\sigma_\textnormal{SIS}$ is the one-dimensional velocity dispersion for the singular isothermal sphere (SIS) density profile. The SIS profile is given by $\rho(r)=\sigma_\textnormal{SIS}/(2\pi Gr^2)$.}

To obtain radial density profiles, a convergence map that sufficiently resolves the relevant strong lensing regime of the lens is created.
We then follow \xuabbr and fit a tenth degree polynomial (in log--log space) to the radial profile.
This polynomial fit is used to derive the lens properties that are under investigation and detailed in section~\ref{sec:formalism}.
We note that another useful characteristic radius used below is the effective radius $r_\textnormal{eff}$, which we define as the projected radius enclosing half of the stellar mass. 
See \appref{app:pdp1} for a description of how $r_\textnormal{eff}$ is calculated.

\subsection{Formalism}
\label{sec:formalism}

We apply the mathematical formalism of \xuabbr to the \textsc{eagle} lenses to assess the bias on $H_0$ and compare the two simulations to one another.
For more details of the calculations presented here, see \xuabbrns.
The main quantities of interest are the average slope between typical radii of lensed images, denoted $s$, and the deviation of the radial profile form a pure power-law (the curvature), denoted $\xi$. 
Following \xuabbrns, we evaluate the convergence at 0.5 and 1.5 times the Einstein radius and denote the radii as $\theta_1$ and $\theta_2$, respectively.
Similarly, we denote the values of the convergence at $\theta_1$ and $\theta_2$ as $\kappa_1$ and $\kappa_2$, respectively.
We can then define
\begin{equation}
s \equiv -\frac{\ln(\kappa_2/\kappa_1)}{\ln(\theta_2/\theta_1)},
\end{equation}
and the curvature
\begin{equation}
\xi \equiv \frac{\kappa(\sqrt{\theta_1\theta_2})}{\sqrt{\kappa_1\kappa_2}}.
\end{equation}

The MSD maps the true $s$ and $\xi$ into ``measured'' values denoted by $s_\lambda$ and $\xi_\lambda$, respectively. The latter two are similarly given by
\begin{equation}
s_\lambda \equiv -\frac{\ln(\kappa_\lambda(\theta_2)/\kappa_\lambda(\theta_1))}{\ln(\theta_2/\theta_1)},
\end{equation}
and
\begin{equation}
\xi_\lambda \equiv \frac{\kappa_\lambda(\sqrt{\theta_1\theta_2})}{\sqrt{\kappa_\lambda(\theta_1)\kappa_\lambda(\theta_2)}}.
\end{equation}
Thus, by using power-law models, we implicitly set $\xi_\lambda=1$, picking out a particular MST; this condition leads to a bias of 
\begin{equation}
\lambda=\frac{\kappa_2+\kappa_1-2\xi\sqrt{\kappa_2\kappa_1}}{\kappa_2+\kappa_1-2\xi\sqrt{\kappa_2\kappa_1}+(\xi^2-1)\kappa_2\kappa_1}.
\end{equation}
Note that if $\xi=1$, the true radial profile $\kappa(\theta)$ is a power-law, and $\lambda=1$.
If $\xi\ne 1$, then there will be a bias, and a fit to observational data would infer a power-law slope that is different from the true slope; we denote this inferred slope by $s_\lambda$.

We also perform a similar set of calculations for the mean convergence within a particular radius, given by 
\begin{equation}
\bar{\kappa}(\theta) = \frac{1}{\pi\theta^2}\int_0^\theta\kappa(\theta')2\pi\theta'\md\theta',
\end{equation}
where $\bar{\kappa}$ can be related to the deflection angle; this can be useful if the deflection angle is expected to follow a power-law.
Analogously to the calculations of the convergence, slope, and curvature, we can define similar quantities for the mean convergence: $\bar{s}$, $\bar{\xi}$, and $\bar{\lambda}$.

Observationally, the only meaningful MST is with respect to the total mass density profile and results in $\lambda>0$ or $\bar{\lambda}>0$.
Radial profiles that require $\bar{\lambda}<0$ have shallow density profiles and/or large curvatures.
Such an MST would result in $s_\lambda<0$; i.e. the density would increase with radius.
The above quantities can be computed for the baryonic and dark matter components separately.
Theoretically, studying the individual components could possibly help discover causes of bias or properties of lenses with the least bias and scatter in $H_0$.
Finally, we note that although these quantities do not take into account any lensing observables, they are still useful indicators of the potential bias and scatter in measurements of the Hubble constant. A more thorough analysis is given in section~\ref{sec:modelling}.

\subsection{Results}
\label{sec:results1}

\begin{table*}
\centering
\caption{Key statistics for the lens populations detailed in section~\ref{sec:comparison}. (See the text for a description of selection criteria and methods.) Columns 2--4 represent a fixed source redshift of $z_s=1.5$ and varying lens redshift. Columns 5--7 represent a fixed lens redshift of $z_d=0.615$ and varying source redshift. Rows 4--13 show, for various source-lens redshift combinations, fractions of meaningful mass-sheet transformations, Einstein radii, and SIS velocity dispersions. Rows 14--17 show total and stellar masses within $R_{200}$, the spherical radius within which the mean density is 200 times the critical density of the Universe, extracted from the raw particle data for the various lens redshifts. The total number of projections includes up to three projections of the same galaxy. A meaningful MST requires that either $\lambda$, $\bar{\lambda}$, or both are positive and nonzero. The minimum Einstein radii correspond to the SIS velocity dispersion cut of $\sigma_\textnormal{SIS}>160\,\mathrm{km}\,\mathrm{s}^{-1}$. \xucftab{1}}
\begin{tabular}{  l || c c c | c c c}
\hline
sample sets & \multicolumn{3}{c|}{$z_s$=1.5} & \multicolumn{3}{c}{$z_d$=0.615} \\
\hline
redshifts & $z_d$=0.183 & $z_d$=0.366 & $z_d$=0.865 & $z_s$=1 & $z_s$=1.5 &$z_s$=3 \\
\hline
total number of projections & 841 & 1074 & 258 & 817 & 1066 & 1186 \\
\hline
meaningful MST for $\kappa$ & 97\% & 96\% & 97\% & 96\% & 98\% & 97\% \\
meaningful MST for $\bar{\kappa}$ & 93\% & 94\% & 97\% & 93\% & 95\% & 96\% \\
meaningful MST for both $\kappa$ and $\bar{\kappa}$ & 92\% & 92\% & 95\% & 92\% & 94\% & 94\% \\
$r_\textnormal{Ein}^\textnormal{min}$ [kpc] & 1.94 & 2.60 & 3.12 & 1.63 & 2.48 & 3.31 \\
$r_\textnormal{Ein}^\textnormal{max}$ [kpc] & 12.61 & 23.35 & 11.54 & 9.69 & 17.42 & 26.98 \\
mean $r_\textnormal{Ein}$ [kpc] & 3.17 & 4.14 & 4.30 & 2.64 & 3.89 & 5.16 \\
median $r_\textnormal{Ein}$ [kpc] & 2.73 & 3.52 & 3.98 & 2.28 & 3.34 & 4.36 \\
standard deviation $\sigma_{r_\textnormal{Ein}}$ [kpc] & 1.29 & 1.79 & 1.20 & 1.03 & 1.63 & 2.23 \\
median $\sigma_{\rm{SIS}}$ [$\mathrm{km}\,\mathrm{s}^{-1}$] & 190.1 & 186.4 & 233.5 & 189.4 & 185.7 & 183.7 \\
stand. dev. $\sigma_{\rm{SIS}}$ [$\mathrm{km}\,\mathrm{s}^{-1}$] & 36.8 & 37.7 & 30.6 & 35.6 & 36.3 & 36.7 \\\hline
median $\log_{10}(M_{200,\rm{tot}}/\rm{M}_{\astrosun}$) & 12.72 & 12.63 & 12.92 & -- & 12.54 & -- \\
stand. dev. $\log_{10}(M_{200,\rm{tot}}/\rm{M}_{\astrosun}$) & 0.45 & 0.43 & 0.38 & -- & 0.40 & -- \\
median $\log_{10}(M_{200,\rm{*}}/\rm{M}_{\astrosun}$) & 11.17 & 11.11 & 11.34 & -- & 11.02 & -- \\
stand. dev. $\log_{10}(M_{200,\rm{*}}/\rm{M}_{\astrosun}$) & 0.33 & 0.33 & 0.30 & -- & 0.32 & -- \\
\hline
\end{tabular}

\label{tab:galstat}
\end{table*}

\begin{figure}
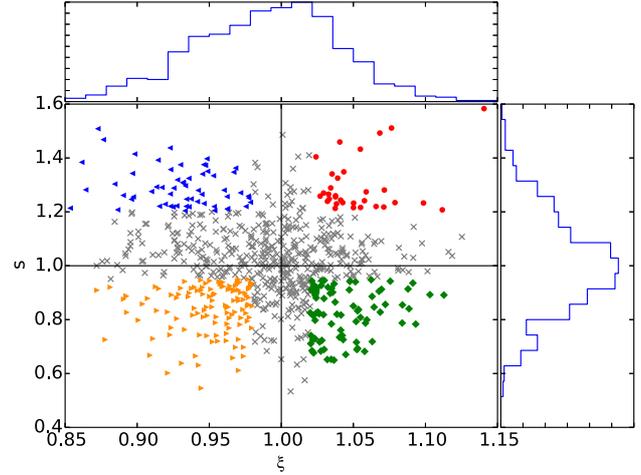

\centering
\includegraphics[width=\linewidth]{{{fig_1.z000p183}.eps}}
\caption{Projected density slope versus curvature parameter. The plotting style and colour-coding separate the lenses into those that are sub- or super-isothermal and concave or convex. They serve as a visual guide for examining subsequent plots. Lenses that lie on the $\xi=1$ line would give an unbiased measurement of $H_0$. \xucf{1}}
\label{fig:fig1}
\end{figure}

\begin{figure}
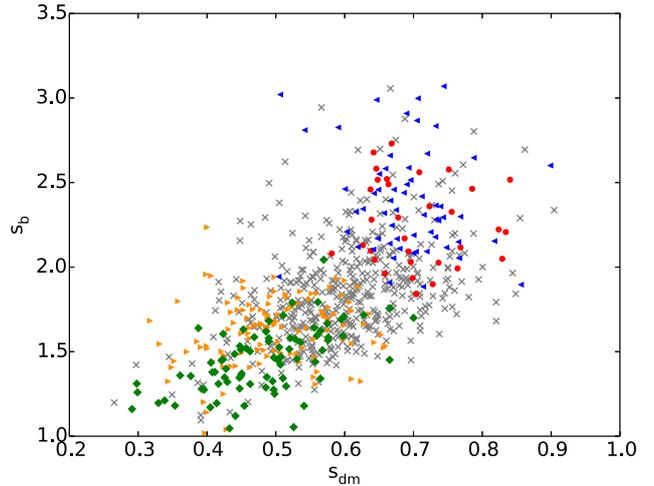

\centering
\includegraphics[width=\linewidth]{{{fig_6.z000p183}.eps}}
\caption{Projected density slope of the baryons versus that of the dark matter. The plotting style and colour-coding are consistent with Fig.~\ref{fig:fig1}. \xucf{6}}
\label{fig:fig6}
\end{figure}

\begin{figure*}
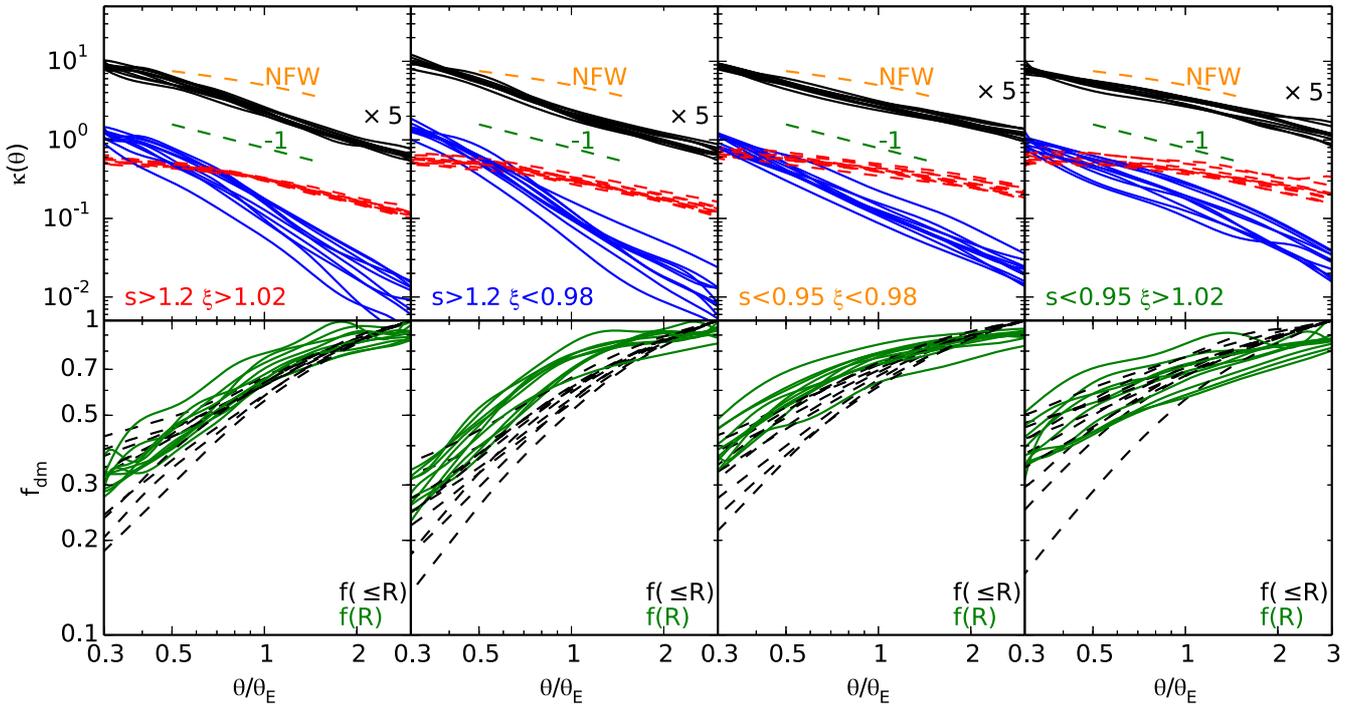

\centering
\includegraphics[width=\linewidth]{{{fig_3.z000p183}.eps}}
\caption{Convergence and dark matter fraction as functions of the radius in units of Einstein radius. Top row: The convergence profile for the baryons (solid blue) and dark matter (dashed red) components. The total convergence is also shown in solid black, but it has been scaled up by a factor of five for clarity. The dashed green lines represents an isothermal slope between 0.5 and 1.5 Einstein radii, while the dashed orange curves represent the projected NFW halo, assuming a scale radius of $10\theta_\textnormal{E}$ \citep{projnfw}. Bottom: The solid green lines represent the projected fractional dark matter density, while the dashed black lines represent the ratio of the enclosed projected dark matter mass to the total projected mass within a given radius. The columns distinguish between the different plotting schemes of Fig.~\ref{fig:fig1} (given in the bottom left of the top row). For each column, we only show results for ten typical lens profiles. \xucf{3}}
\label{fig:fig3}
\end{figure*}

\begin{figure}
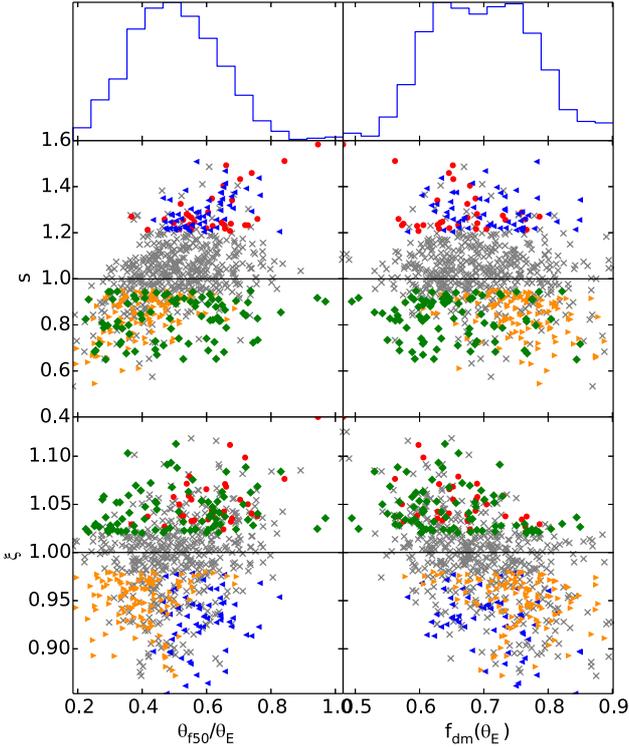

\centering
\includegraphics[width=\linewidth]{{{fig_7.z000p183}.eps}}
\caption{Projected density slope $s$ and curvature parameter $\xi$ versus the equivalence radius $\theta_\textnormal{f50}$ (in units of the Einstein radius) and local, projected fractional dark matter density at the Einstein radius. $\theta_\textnormal{f50}$ is defined as the radius at which the local, projected dark matter and baryon fractions are equal. The plotting style and colour-coding are consistent with Fig.~\ref{fig:fig1}. \xucf{7}}
\label{fig:fig7}
\end{figure}
\begin{figure*}
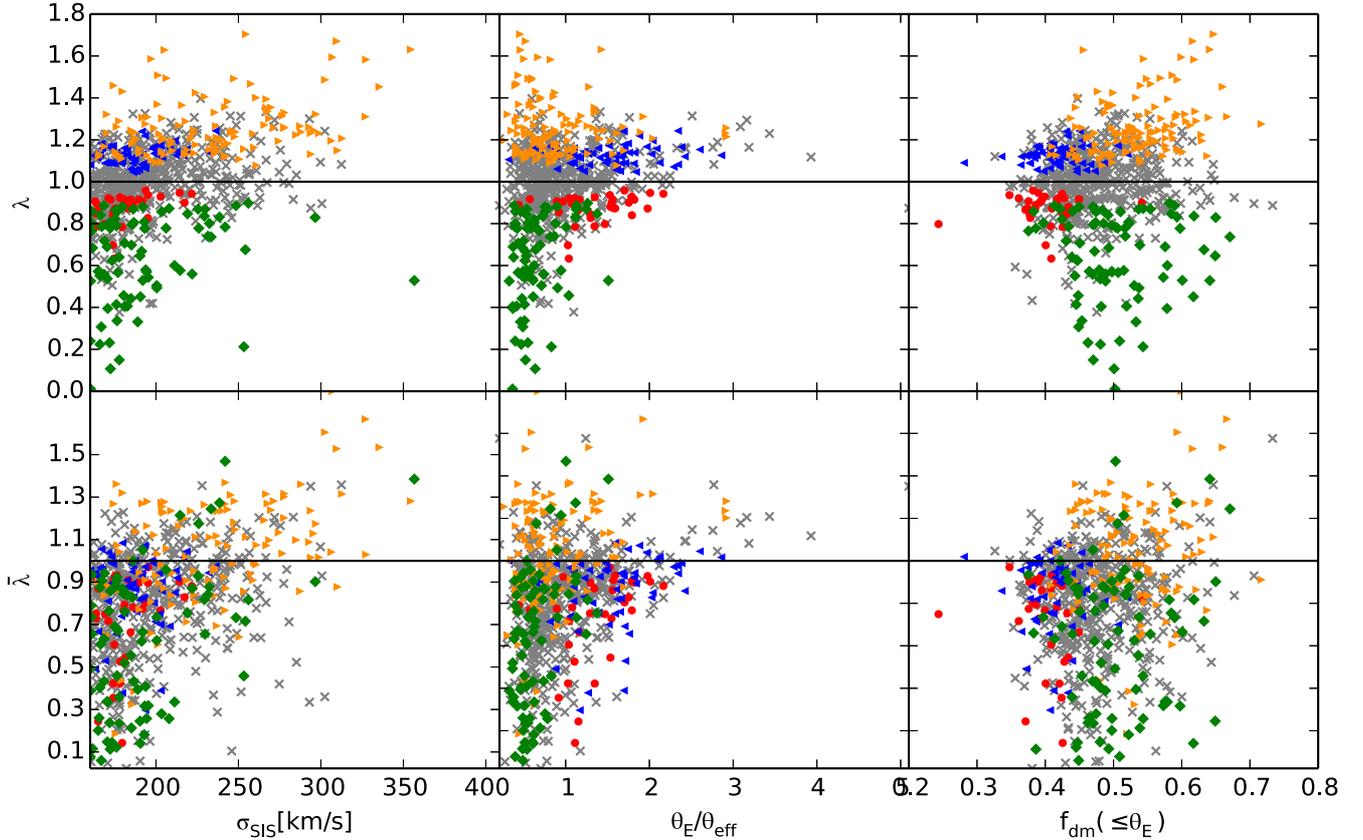

\centering
\includegraphics[width=\linewidth]{{{fig_8.z000p183}.eps}}
\caption{Biases on $H_0$ versus the velocity dispersion calculated from the Einstein radius assuming a singular isothermal model, the Einstein radius normalized in units of the effective radius, and the fraction of dark matter within the Einstein radius. The top (bottom) row shows the bias, assuming a power-law profile for the density slope (deflection angle). The plotting style and colour-coding are consistent with Fig.~\ref{fig:fig1}. \xucf{8}}
\label{fig:fig8}
\end{figure*}

\begin{figure}
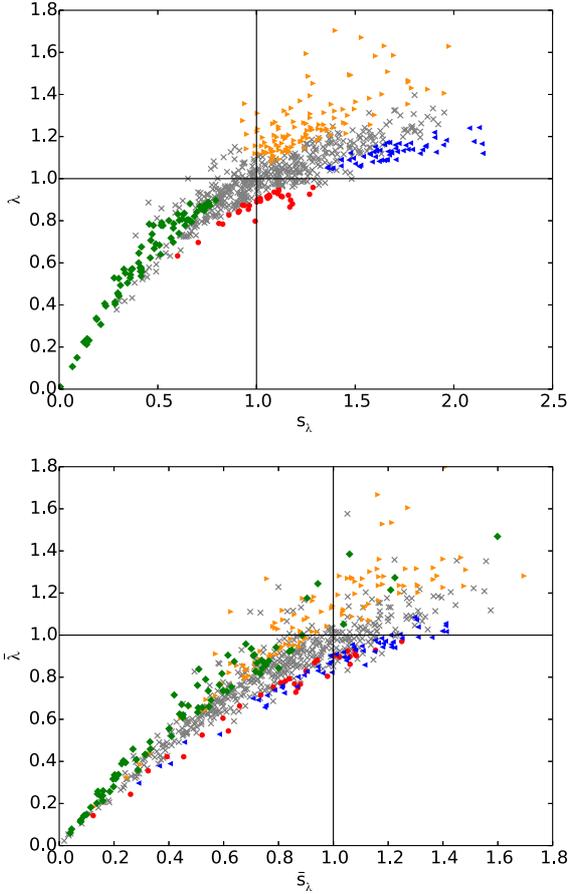

\centering
\begin{tabular}{ c }
\includegraphics[width=0.9\linewidth]{{{fig_9a.z000p183}.eps}} \\
\includegraphics[width=0.9\linewidth]{{{fig_9b.z000p183}.eps}}
\end{tabular}
\caption{Bias on $H_0$ versus projected density slope. Top: Assumes power-law profile in density. Bottom: Assumes power-law profile in deflection angle. The plotting style and colour-coding are consistent with Fig.~\ref{fig:fig1}. \xucf{9}}
\label{fig:fig9}
\end{figure}

\begin{table*}
\centering
\caption{Statistics for lenses with MST-transformed density slopes and/or mean convergence profiles near unity. \xucftab{2}}
\begin{tabular}{  l || c c c | c c c}
\hline
Sample sets & \multicolumn{3}{c}{$z_s=1.5$} & \multicolumn{3}{c}{$z_d=0.615$} \\
Redshifts & $z_d=0.183$ & $z_d=0.366$ & $z_d=0.865$ & $z_s=1.0$ & $z_s=1.5$ & $z_s=3.0$ \\
\hline
\rule{0pt}{1ex}   \\
\multicolumn{7}{c}{Subsample I: {$s_\lambda\in [0.9,1.1]$}} \\
Number of galaxy projections & 168 & 217 & 37 & 131 & 189 & 220 \\
Mean $\lambda$ & 1.01 & 1.03 & 1.08 & 1.00 & 1.03 & 1.02 \\
Median $\lambda$ & 1.00 & 1.01 & 1.07 & 0.99 & 1.01 & 1.00 \\
Standard deviation of $\lambda$ & 0.10 & 0.10 & 0.09 & 0.10 & 0.11 & 0.09 \\
\rule{0pt}{1ex}   \\
\multicolumn{7}{c}{Subsample II: {$\bar{s}_\lambda\in [0.9,1.1]$}} \\
Number of galaxy projections & 156 & 220 & 56 & 183 & 233 & 259 \\
Mean $\bar{\lambda}$ & 0.98 & 1.05 & 1.05 & 0.97 & 1.03 & 1.04 \\
Median $\bar{\lambda}$ & 0.96 & 1.03 & 1.04 & 0.96 & 1.01 & 1.02 \\
Standard deviation of $\bar{\lambda}$ & 0.12 & 0.13 & 0.13 & 0.13 & 0.14 & 0.11 \\
\rule{0pt}{1ex}   \\
\multicolumn{7}{c}{Subsample III: {$s_\lambda\in [0.9,1.1]$ and $\bar{s}_\lambda\in [0.9,1.1]$}} \\
Number of galaxy projections & 30 & 54 & 11 & 32 & 35 & 59 \\
Mean $\lambda$ & 0.99 & 1.05 & 1.06 & 0.99 & 1.05 & 1.02 \\
Median $\lambda$ & 0.99 & 1.04 & 1.04 & 0.98 & 1.03 & 1.01 \\
Standard deviation of $\lambda$ & 0.09 & 0.09 & 0.08 & 0.10 & 0.08 & 0.08 \\
Mean $\bar{\lambda}$ & 0.99 & 1.06 & 1.08 & 0.97 & 1.04 & 1.03 \\
Median $\bar{\lambda}$ & 0.99 & 1.04 & 1.04 & 0.95 & 1.00 & 1.01 \\
Standard deviation of $\bar{\lambda}$ & 0.10 & 0.13 & 0.10 & 0.11 & 0.11 & 0.09 \\
\end{tabular}
\label{tab:ssl}
\end{table*}

\tabref{tab:galstat} presents key statistics about the lens populations.
Qualitatively, the properties of the lenses in the three lowest (lens) redshift bins are consistent with those of \xuabbrns.
The number of lenses, as well as the mean and median Einstein radii, show similar trends as a function of the lens redshift.
Depending on whether all lenses are selected or only those which produce meaningful MSTs ($\lambda>0$, $\bar{\lambda}>0$, or both), the number of lenses can vary significantly.
The Illustris lenses produce many more lens projections, but the \textsc{eagle} lenses have a higher rate of meaningful MSTs.
The \textsc{eagle} lenses also have smaller mean and median Einstein radii, which are 70--85\% the size of Illustris lenses.
Unsurprisingly, the standard deviations of the Einstein radii are also smaller.
For the three lowest lens redshifts, these effects persist after accounting for the slight differences in redshift; we do not extrapolate the results of \xuabbr to compare the highest redshift bin.

The reason for this difference could be due to the larger size of Illustris galaxies, as they do not reproduce the observed galaxy size--mass relation \citep{illsizemass}.
Assuming, for simplicity, spherically symmetric density profiles, the Einstein radius depends not only on how centrally concentrated the radial profile is but also on the form of the profile.
In other words, a more centrally concentrated galaxy does not necessarily produce a smaller or larger Einstein radius, compared to a more extended galaxy.
Comparing the detailed profiles of lenses in these simulations to one another and their relation to the lens properties is beyond the scope of this work.
Nevertheless, we note that the largest discrepancy occurs for the highest redshift bin, $z_d=0.865$, where we find a drastically smaller number of lenses and much larger mean and median Einstein radii, compared to the other lens redshift bins. 

We further attempt to compare the two simulations by using \textsc{eagle} lenses to reproduce several key figures in \xuabbrns; for consistency, the results presented in this section are only for the $z_d=0.183$ and $z_s=1.5$ combinations.
Fig.~\ref{fig:fig1} shows two important properties of the lenses: the density slope and the curvature at a radius between the two image positions.
It also defines the separation of the variously coloured lenses based on the slope and curvature parameter, which show no discernible difference from \xuabbrns.
The lenses are separated into those which are sub-isothermal ($s<0.95$), super-isothermal ($s>1.2$), convex downwards ($\xi<0.98$), and concave upwards ($\xi>1.02$).
The lenses are centered around radial profiles that are isothermal with no curvature, but there is significant scatter in both parameters.
This is consistent with previous results that have shown gravitational lenses to be, on average, isothermal \citep[see e.g.][]{avgiso}. 

Compared to those that only contain dark matter, simulations that include baryons show cuspier central profiles, which is likely due to mass-dependent effects of the baryons, such as adiabatic contraction \citep{schaller,barydm}.
Fig.~\ref{fig:fig6} decomposes the lens slope into the constituent dark matter and baryonic components.
It confirms the super-isothermal slopes of the baryons and sub-isothermal dark matter profiles, and it shows a clear correlation between the two, i.e. lenses that have steeper profiles ($s>1$) also have steeper baryonic density slopes ($s_b\gtrsim2$).

The top row of Fig.~\ref{fig:fig3} shows the density profiles for ten typical galaxies from each of the four categories outlined in Fig.~\ref{fig:fig1}. 
The baryonic (solid blue line) and dark matter (dashed red line) components, as well as their sums (solid black), are shown separately.
It is clear that the baryons have a steeper density slope and are more centrally concentrated; the dark matter, on the other hand, begins to dominate somewhere between the 0.3 and 1 Einstein radii.
The bottom row of Fig.~\ref{fig:fig3} shows the cumulative (dashed black line) and local dark matter fractions (solid green line) for the same 10 galaxies shown in each panel of the top row.
Due to the steepness of the baryonic density profile in the left two columns ($s>1.2$), the dark matter fraction rises quickly inside the Einstein radius.
On the other hand, the right two columns, where $s<0.95$, show shallower dark matter fraction curves.
Compared to \xuabbr, there are two main differences.
For lenses with $\xi>1.02$, the radius at which the dark matter begins to dominate is typically smaller.
This transition appears to occur near the Einstein radius for the Illustris sample, but it occurs at $\sim 0.5 r_\textnormal{Ein}$, or at even smaller radii, for \textsc{eagle} lenses.
Another notable difference is that (for \textsc{eagle}) the curvature in lenses with $s<0.95$ is, qualitatively, much less pronounced in both the density profiles of the baryonic and total matter densities.
This could be, however, an effect of the difference in physical scales being probed by the two simulations.

Fig.~\ref{fig:fig7} shows the density slope and curvature as functions of the equivalence radius $\theta_\textnormal{f50}$ (the radius at which the local, projected density of baryons equals that of the dark matter) and the local, projected fraction of dark matter at the Einstein radius.
The difference in size between \textsc{eagle} and Illustris lenses is evident here.
The corresponding figure in \xuabbr shows that the density of baryons can dominate in many cases up to 1.5 times the Einstein radius.
The equivalence radius for the \textsc{eagle} lenses, on the other hand, can be as large as the Einstein radius but is typically smaller with a median value of 0.50 times the Einstein radius.
Aside from this, the only other notable difference stems from the sub-isothermal lenses.
In this group, \xuabbr show a clear separation between the $\xi<0.98$ (orange) and $\xi>1.02$ (green) groups.
The \textsc{eagle} lenses show a significant overlap in these groups, which may be attributable to the milder curvatures seen in Fig.~\ref{fig:fig3}.

To assess any possible bias on the Hubble constant, $\lambda$ and $\bar{\lambda}$ must be evaluated.
In Fig.~\ref{fig:fig8} we show the distributions of these multiplicative biases as functions of the velocity dispersion, the Einstein radius (normalized by the effective radius), and the cumulative dark matter fraction.
Like \xuabbrns, we find a correlation of the bias with $\sigma_\textnormal{SIS}$.
Lenses with smaller velocity dispersions ($<200\,\mathrm{km}\,\mathrm{s}^{-1}$) tend to be heavily biased with values of $\lambda$ and $\bar{\lambda}$ that approach zero.
Similarly, we find that lenses with larger velocity dispersion are less biased but still have significant scatter.
For lenses with $\sigma_\textnormal{SIS}>200\,\mathrm{km}\,\mathrm{s}^{-1}$, the median bias is 1.10 with a standard deviation of 0.48, or 44\%.
Additionally, there are several differences between the two lens samples that can be seen here.
There are fewer \textsc{eagle} lenses with $\sigma_\textnormal{SIS}>350\,\mathrm{km}\,\mathrm{s}^{-1}$, and the distributions of $\theta_\textnormal{E}/\theta_\textnormal{eff}$ extend several factors higher.
Consequentially, the larger Einstein radii lead to larger fractions of dark matter within the Einstein radius itself (as seen in the rightmost column).
However, these differences are likely linked to the previously seen difference in lens galaxy sizes between the two simulations. 

Fig.~\ref{fig:fig9} shows a clear correlation of the bias with the density slope after the MST has been applied.
It is clear from both panels that lenses near isothermality lead to the smallest bias and that the scatter is significantly reduced in this regime to $<20\%$.
It is important to note that the transformed density slopes, $s_\lambda$ and $\bar{s}_\lambda$, are what one would infer from observational data. 
These findings are in good qualitative agreement with what \xuabbr find and motivate the authors to extract only those lenses with ``measured'' density slopes between 0.9 and 1.1. 
They find that the bias on $H_0$ from these subsamples can be significantly reduced to less than 5\%.
Similarly, we focus on the the subsamples of \textsc{eagle} lenses for which $s_\lambda$ and $\bar{s}_\lambda$ fall into the same range of density slopes.
In \tabref{tab:ssl} we show key statistics for the various lens and source redshift combinations.
Like \xuabbrns, we find that selecting lenses with $s_\lambda$ or $\bar{s}_\lambda$ near unity significantly reduces the bias on $H_0$.
The standard deviations of $\lambda$ or $\bar{\lambda}$ are also reduced to $\sim 10\%$.
\xuabbr see a further reduction in scatter to 5\% when requiring that both $s_\lambda$ and $\bar{s}_\lambda$ be near unity.
When a similar selection is applied to the \textsc{eagle} lenses, the number of lenses in the selection set is reduced but the scatter does not significantly change. 

The properties of the radial profiles of lenses from the \textsc{eagle} simulation are broadly consistent with those of \xuabbrns.
We suspect that the primary differences arise due to the differences in the size of galaxies between the two simulations.
\textsc{eagle} galaxies reproduce the observed galaxy size--mass relation, whereas Illustris galaxies do not.
This has a significant effect on the measured effective radii and Einstein radii of the galaxies, which, in turn, affect several quantities, such as the equivalence radius.
However, the implications for measurements of the Hubble constant remain the same.
By selecting galaxies near isothermality, the bias on $H_0$ can be reduced to $\sim5\%$ and the scatter can be reduced to $\sim10\%$.

\section{Joint lens and dynamical modelling}
\label{sec:modelling}

The results in the previous section are promising and suggest that by selecting galaxies with large velocity dispersions and/or near-isothermal density slopes, the bias and scatter on $H_0$ can be minimized.
However, the underlying method in the analysis was to transform the observed density of lens galaxies in the simulation so that the curvature parameter became $\xi_\lambda=1$.
In other words, the transformed convergence at 0.5, 1.5, and $\sqrt{0.5\times1.5}$ Einstein radii all lie on a line in log--log space; i.e., the transformed convergence is a power-law.
Here, we attempt to assess what bias remains on $H_0$ after taking into account lensing observables for quasar images and kinematic information.
In section~\ref{sec:adv} we also take into account emission from the quasar host galaxy and the effects of the lens environment.

In this section, we perform a joint lensing and dynamics analysis for a subset of the lenses presented in the previous sections.
Starting from a three-dimensional gravitational density, model predictions can be made in a self-consistent way under certain assumptions discussed in section~\ref{sec:kinematics}.
In other words, model predictions for both the lensing observables and the velocity dispersion measurements can be obtained from any given three-dimensional density distribution model.
Thirteen parameters, given in \tabref{tab:modelparams1}, define the density model and include the position, mass, density slope, axes ratios, ellipsoid orientation, viewing angle (or equivalently the position angle of the projected ellipticity), effective radius, core radius, and truncation radius. One additional parameter is the Hubble parameter, which is, of course, free to vary.

We use a hybrid code framework to do the modelling and sample the parameter space.
The actual model fitting is performed analytically (as opposed to numerically), and the code used is publicly available.\footnote{https://github.com/tagoreas/Lensing-code}
However, modelling hundreds of lenses and thoroughly exploring the parameter space can be computationally demanding.
We therefore use the Python module \textsc{emcee} \citep{emcee} to run a Monte Carlo Markov Chain analysis.
The code uses an affine-invariant sampling method to achieve fast convergence.
300 walkers suffice to fully explore the parameter space, and after a burn-in phase of 150 steps, we take an additional 150 steps to compute the posterior probability distribution.

\subsection{Simulating lenses and extracting kinematic observables}
\label{sec:dataextract}

We wish to extract lensing and kinematic observables from the simulation.
The lensing data we wish to extract include image positions and time delays.
Image fluxes or magnifications are related to second derivatives of the lens potential; because of the limited mass resolution, calculating the second derivatives can be unreliable, especially for less massive galaxies and images near a critical curve.
We, thus, do not use image fluxes as observational constraints.
Lastly, we also extract aperture velocity dispersions within $r_{\textnormal{eff}}/8$ to further constrain the density slope.

A number of practical considerations are made in transforming the particle data into predictions of what would be observed in reality; for a detailed description see \appref{app:pdp2}.
For each lens, we calculate potential, convergence, and deflection maps; the mass within the Einstein radius; the three-dimensional and two-dimensional axis ratios and orientations; and the velocity dispersion within $r_{\textnormal{eff}}/8$.
These properties give all the information needed to generate positions of lensed images, time delays, and an aperture velocity dispersion.
Some of the galaxy properties, such as the three-dimensional orientation of the galaxy, are not directly observable but are directly modelled (see section~\ref{sec:modelling}), and we compare the fitted model parameters to those extracted from the simulation.

\subsection{Lens models}
\label{sec:lensmodels}

\begin{table*}
\centering
\caption{Model parameters and priors for modelling lensing and kinematic observables. The priors are not truly hard, uniform priors, but outside the range specified, a steep penalty function is imposed.}
\begin{tabular}{ c c l }
\hline
Symbol              &  Uniform priors & Description \\
\hline
$x$ & -500--500 mas & Offset of lens position in $x$-direction, relative to minimum of potential \\
$y$ & -500--500 mas & Offset of lens position in $y$-direction, relative to minimum of potential \\
$\log_{10}(\rm{M}_{\textnormal{r}_\textnormal{Ein}}/\rm{M}_{\astrosun})$ & 7--15 & Projected mass within Einstein radius \\
$\gamma$ & 1.5--2.5 & Three-dimensional density slope \\
$q$ & 0.2--1 & Intermediate axis ratio, relative to major axis \\
$p$ & 0.2--$q$ & Minor axis ratio, relative to major axis \\
$\theta_\textnormal{xy}$ & $0$--$90^\circ$ & Viewing angle in $x$--$y$ plane from $+x$-axis  \\
$\phi_\textnormal{z}$ & $0$--$90^\circ$ & Viewing angle from $+z$-axis \\
$\theta_\textnormal{PA}$ & $-90$--$90^\circ$ & Position angle of ellipticity \\
$H_0$ & 10--150 $\mathrm{km}\,\mathrm{s}^{-1}\mathrm{Mpc}^{-1}$ & Hubble constant \\
$r_\textnormal{eff}$ & 0.05--10 $\mathrm{arcsec}$ & Two-dimensional (circularized) effective radius \\
$r_c$ & 0.01--1 $r_\textnormal{eff}$ & Three-dimensional core radius \\
$r_t$ & 10--100 $r_\textnormal{eff}$ & Three-dimensional truncation radius \\
\hline
\end{tabular}
\label{tab:modelparams1}
\end{table*}

\begin{table*}
\centering
\caption{Observational constraints and assumed (Gaussian) uncertainties. For fractional uncertainties, a minimum uncertainty is required.}
\begin{tabular}{ c c c l }
\hline
Symbol              &  Uncertainty & Minimum & Description \\
\hline
$g_x$ & 100 mas & -- & $x$-coordinate of lens galaxy (assumed to coincide with peak in potential) \\
$g_y$ & 100 mas &  -- & $y$-coordinate of lens galaxy (assumed to coincide with peak in potential) \\
$r_\textnormal{eff}$ & 10\% & 50 mas & Half-mass radius \\
$e$ & 0.2 & -- &  ellipticity (from S\'{e}rsic fit to surface brightness) \\
PA & $10^\circ$ & -- &  position angle (from S\'{e}rsic fit to surface brightness) \\
$i_x$ & 50 mas & -- & $x$-coordinate of lensed image \\
$i_y$ & 50 mas &  -- & $y$-coordinate of lensed image \\
$\Delta t$ & 3\% & 0.5 days & time delay between images \\
$\sigma_\textnormal{los}$ & $10\,\mathrm{km}\,\mathrm{s}^{-1}$ & -- & line-of-sight velocity disperion within $r_\textnormal{eff}/8$ \\
\hline
\end{tabular}
\label{tab:modelparams2}
\end{table*}

\begin{figure}
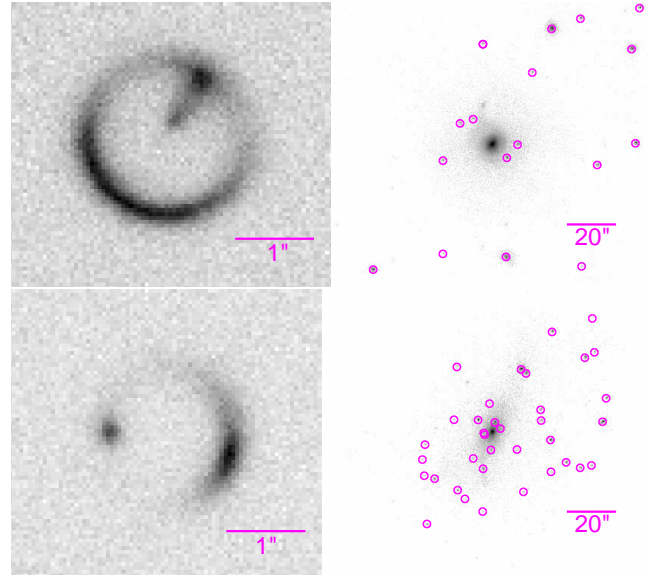

\centering
\includegraphics[width=\linewidth]{{{simlenses}.eps}}
\caption{Simulated lensed surface brightness distribution from the quasar host galaxy (left) and larger-scale lens environment with detected satellite galaxies encircled (right). The colour-scale is linear for the simulated host galaxy but logarithmic for the lens environment so that the satellites are visible. The top row corresponds to one of two lenses, $\mathscr{L}_1$, modelled in section~\ref{sec:adv}. Similarly the bottom row corresponds to the other lens, $\mathscr{L}_2$. See the text for details about the host galaxy simulation and satellite detection.}
\label{fig:extsb}
\end{figure}

As mentioned previously, the lenses are modelled as softened, truncated, triaxial power-law ellipsoids with a three-dimensional density given by
\begin{equation}
\rho(r) = \rho_0\big[(r_c^2+r^2)^{-\gamma/2} - (r_t^2+r^2)^{-\gamma/2}\big],
\label{eq:density}
\end{equation}
where 
\begin{equation}
r^2 = \Big(\frac{x}{r_{\textnormal{3d}}}\Big)^2+\Big(\frac{y/q}{r_{\textnormal{3d}}}\Big)^2+\Big(\frac{z/p}{r_{\textnormal{3d}}}\Big)^2
\end{equation}
and $r_{\textnormal{3d}}$ is the three-dimensional effective radius measured along the major axis, $\rho_0$ is the density at $r\sim r_{\textnormal{3d}}$, $r_c$ and $r_t$ are the core and truncation radii, respectively, in units of $r_{\textnormal{3d}}$ ($r_c<r_t$), $\gamma$ is the density slope, $(x,y,z)$ are Cartesian coordinates along the principal axes of the ellipsoidal density distribution, and $1\ge{q}\ge{p}>0$.
This density profile is especially useful because it ensures a finite mass given, for the spherical case, by 
\begin{equation}
M_\textnormal{inf} = \pi^{3/2}\rho_0(r_c^{3-\gamma}-r_t^{3-\gamma})\Gamma\Big(\frac{\gamma-3}{2}\Big)/\Gamma\Big(\frac{\gamma}{2}\Big),
\end{equation}
where $\Gamma$ is the gamma function.

The rapidly falling density outside the truncation radius and the lack of a central cusp are desirable for numerical stability in the dynamical modelling (see section~\ref{sec:kinematics}).
Additionally, because numerical simulations cannot resolve the innermost regions of galaxies, a profile with a core radius is practical, especially when modelling the host galaxy, which can produce central images as seen in Fig.~\ref{fig:extsb}.
The near power-law behaviour close to the typical locations of lensed images is desirable, since power-laws are commonly used in the literature for lens modelling.

In total, there are 13 free parameters in the model, given in \tabref{tab:modelparams1}, that are used to derive the lensing and velocity dispersion model predictions.
For a given set of viewing angles, $\theta_{xy}$ and $\phi_{z}$, and position angle $\theta_\textnormal{PA}$ the projected surface density can be calculated analytically.
Each term on the right hand side of \eqref{eq:density} has a corresponding convergence of the form
\begin{equation}
\kappa(r,\theta) = \kappa_0 \Bigg[1+\Big(\frac{r}{r_0}\Big)^2\Big(1+\epsilon \cos\big[2(\theta-\theta_\textnormal{PA})\big]\Big)\Bigg]^{-\frac{\gamma-1}{2}},
\label{eq:kappa}
\end{equation}
where $\kappa_0$, $r_0$, and $\epsilon$ depend on the particular choice of model parameters \citep{chae1}.
We follow the methodology of \citet{chae1} and \citet{chae2}, who find fast-converging series solutions for calculating deflections, magnifications, and time delays from \eqref{eq:kappa}.

\subsection{Kinematics}
\label{sec:kinematics}

Given a two-dimensional lensing potential, there are, in general, a number of ways to make a prediction of the velocity dispersion within an aperture.
The simplest analytic approach is to assume that the de-projected density is spherically symmetric and to solve the spherical Jeans equation.
Although this method is quick, it may not be physically well-motivated as galaxies are rarely spherical. 
Another possibility includes using Jeans axisymmetric models \citep{JAM}, which use only the surface brightness distribution of the lens to model the galaxy kinematics \citep{JAMlens}.

If non-spherical density models are considered, a two-dimensional lensing potential does not uniquely determine the dynamics of the galaxy.
For example, a circularly-symmetric lens potential can be consistent with a wide range of velocity dispersions, depending on whether the de-projected density is prolate, spherical, or oblate.  
For this purpose, \citet{cauldron} have developed and tested the \textsc{cauldron} code, which uses axisymmetric models to quickly predict velocity dispersions.

However, galaxies and their halos are generally triaxial \citep[see e.g.][]{triaxial}.
In order to most realistically model their three-dimensional shapes, we rely on the Schwarzschild method, which is an approach to studying the orbits of particles in a gravitational potential \citep{smile}.
Typically, this method numerically follows the trajectories of a large ensemble of particles in the potential. 
The positions and velocities of the particles are tracked, but the density, computed over a number of grid cells, is also computed.
Then, each particle is weighted in such a way that the grid-computed density of the particles and the potential used to generate the trajectories are related to one another via the Poisson equation.

We use the publicly-available Schwarzschild orbit modelling code \textsc{smile} \citep{smile}.
\textsc{smile} allows us to not only track the particles, but it also creates N-body snapshots of the particles.
Given a mass density corresponding to \eqref{eq:density}, we can make predictions for the positions and velocities of the particles.
For creating N-body snapshots, we are only interested in those particles corresponding to the stars and not to the dark matter.
We accomplish this by creating a two-component mass model within the \textsc{smile} framework. 
The first component, given by \eqref{eq:density}, contains 99.999\% of the mass of the system and is the power-law component.
The second component, which contains negligible mass, is chosen to reproduce the deprojected S\'{e}rsic profile (as described in the \textsc{smile} manual), and its axes ratios, orientation, and scale radius are identical to those of the first component.
The N-body snapshots created only contain the weighted orbits of particles that reproduce the second stellar component and, thus, can be used to make model predictions for the central velocity dispersions.
Each snapshot contains $10^6$ particles and was created from $10^4$ orbits, each sampled 100 times.
\textsc{smile} ranks the quality of each model as poor, fair, or good for a number of criteria, such as its numerical feasibility and its uniformity of particle weights.
We only keep snapshots that \textsc{smile} reports as being fair or good across all criteria.
Models are poor only for the most extreme model parameters, such as those with intermediate and minor axes ratios near 0.2.

We note that although the orbit modelling can take into account other properties of galaxies, such as net rotation of the galaxy, velocity anisotropy, and the presence of a massive black hole, we do not include these possibilities here.

It is also worth noting that for a given velocity dispersion measurement, there is a large range of model parameters that can reproduce the measurement.
For example, both prolate and oblate ellipsoids might be able to reproduce the measurement within the same aperture, but they will likely have significantly different masses within that same aperture.
Lensing will be sensitive to this mass difference.
Our aim is to assess to what degree the addition of kinematic constraints will help break degeneracies already present in the lensing observables.

Unfortunately, simulating the particle orbits is a time-consuming process, and so we tabulate velocity dispersions on a seven-dimensional grid of model parameters and interpolate for any given set of parameters.
Briefly, for each pair of viewing angles $\theta_{xy}$ and $\phi_{z}$ on the grid, we interpolate between $\gamma$, $q$, $p$, $r_c$, and $r_t$ using non-localized radial basis functions, minimizing interpolation errors.
This is then followed by bilinear interpolation in $\theta_{xy}$ and $\phi_{z}$.
For a more detailed description of the orbit modelling and details about the interpolation method and errors, see \appref{app:mi}.


\subsection{Lens sample}
\label{sec:lenssample2}

\begin{figure}
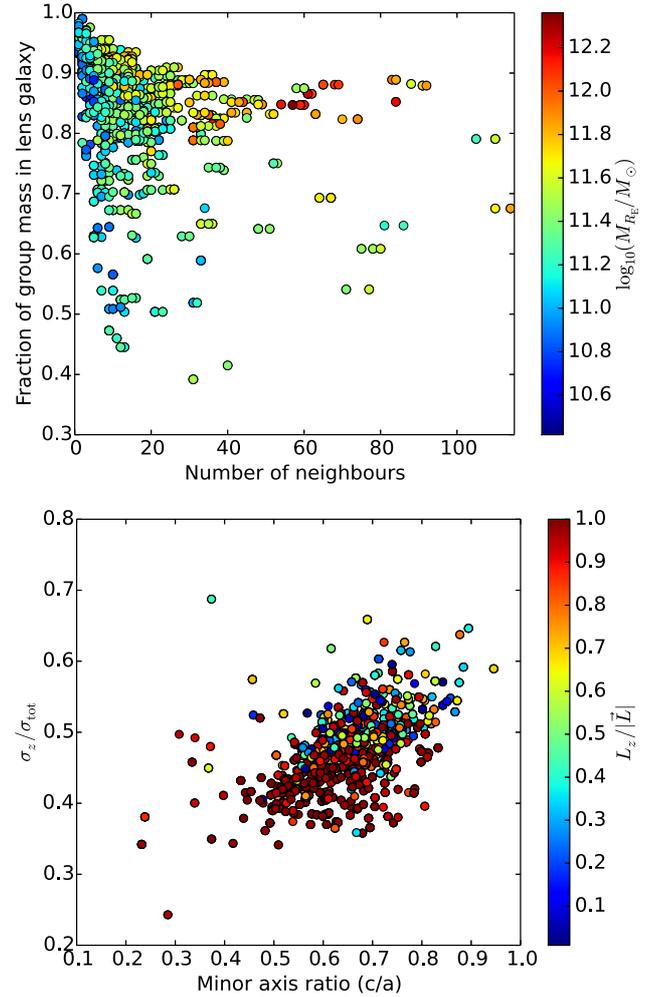

\centering
\begin{tabular}{ c }
\includegraphics[width=\linewidth]{{{neighfrac}.eps}} \\
\includegraphics[width=\linewidth]{{{cLz}.eps}}
\end{tabular}
\caption{Top: Ratio of mass in lens galaxy to friends-of-friend group mass versus the number of significant neighbours a lens has. The colour scale denotes the projected mass within the Einstein radius. Bottom: Fractional velocity dispersion along the minor axis as a function of the minor axis ratio. The normalized component of the angular momentum vector along the minor axis is colour-coded as well. See the text for a description of how neighbours are identified and how these quantities are calculated.}
\label{fig:neighfrac}
\end{figure}

Due to a number of reasons, not all the galaxies that met the selection criteria described in section~\ref{sec:lens_selection} also pass the requirements detailed in \appref{app:pdp2}.
The majority of lens candidates are not relaxed, isolated systems.
By visual inspection, many of them appear to be in group environments with multiple nearby (within three Einstein radii) companions and some are merging or recently merged systems.
We have also examined the \textsc{subfind} catalogues to quantify the environments of the \textsc{eagle} lenses. 
The effect of a particular satellite on deflections and time delays will depend on its mass, its distance from lensed images, and possibly its structure \citep[see e.g.][]{millidefl,millitdel}.
Although there are many possible characterizations, we choose to count, for each lens, the number of galaxies within three different distance bins and three different minimum threshold masses: $5\e{9} \rm{M}_{\astrosun}$ within 50 kpc, $1\e{10} \rm{M}_{\astrosun}$ from 50 kpc to 100 kpc, and $5\e{10} \rm{M}_{\astrosun}$ from 100 kpc to $R_{200}$ (if $R_{200}>100\,\textnormal{kpc}$).
These choices ensure that the satellite galaxies are capable of having a significant effect on the lens modelling.
In Fig.~\ref{fig:neighfrac} we quote the number of neighbours a lens has as the sum over all bins and compare this with the ratio of mass in the lens galaxy to the total friends-of-friends mass.
Because of the scale-dependent selection criterion for identifying neighbours, less massive lenses will naturally have fewer neighbours.
Nevertheless, the figure suggests that lenses with high mass fractions can have various masses but generally have fewer neighbours.
A dense group environment does not automatically make a lens a poor candidate for inferring cosmological parameters, but it does make the modelling more difficult, which could introduce biases.

There are also many disk-like galaxies that qualify as lenses.
These could present a problem because the Schwarzschild orbit modelling used does not take into account galaxy rotation, leading to a bias on the Hubble constant that is dependent on the particular viewing angle for a galaxy.
To try and estimate the number of rapidly rotating objects, we follow a similar procedure to that outlined in \citet{cLz} and examine, for each galaxy's stellar component, the minor-to-major axis ratio $c/a$ and the velocity dispersion along the minor axis $\sigma_z/\sigma_\textnormal{tot}$.
We also calculate the component of the ratio of the angular momentum about the minor axis to the total angular momentum vector $L_z/|\vec{L}|$.
Details of how we compute these quantities can be found in \appref{app:fastrotators}, and the distribution of these quantities can be seen in Fig.~\ref{fig:neighfrac}.
We identify fast rotators as objects with $\sigma_z/\sigma_\textnormal{tot}<0.5$ and $L_z/|\vec{L}|>0.9$.
We do not include the axis ratio as a discriminator because it can depend strongly on the aperture radius chosen for measurement, and the presence of a central bulge can strongly influence the ratio as well.
1061 out of 2195 projected lenses\footnote{The same galaxy may be included multiple times if different projections produce lenses or if a projection produces multiple source--lens configurations (e.g., cusp, double).}, corresponding to 249 out of 533 unique galaxies, satisfy these criteria.
However, not all of them are actually fast rotators; some may be merging systems or have very little angular momentum in the first place.
Regardless of whether or not a galaxy is likely to be a rotating disk or merging system, we include in this analysis all lenses for which lensing observables could be reliably derived from the simulation.

\subsection{Results}
\label{sec:results2}

\subsubsection{Joint lens and dynamical modelling}
\label{sec:rr1}

\begin{figure*}
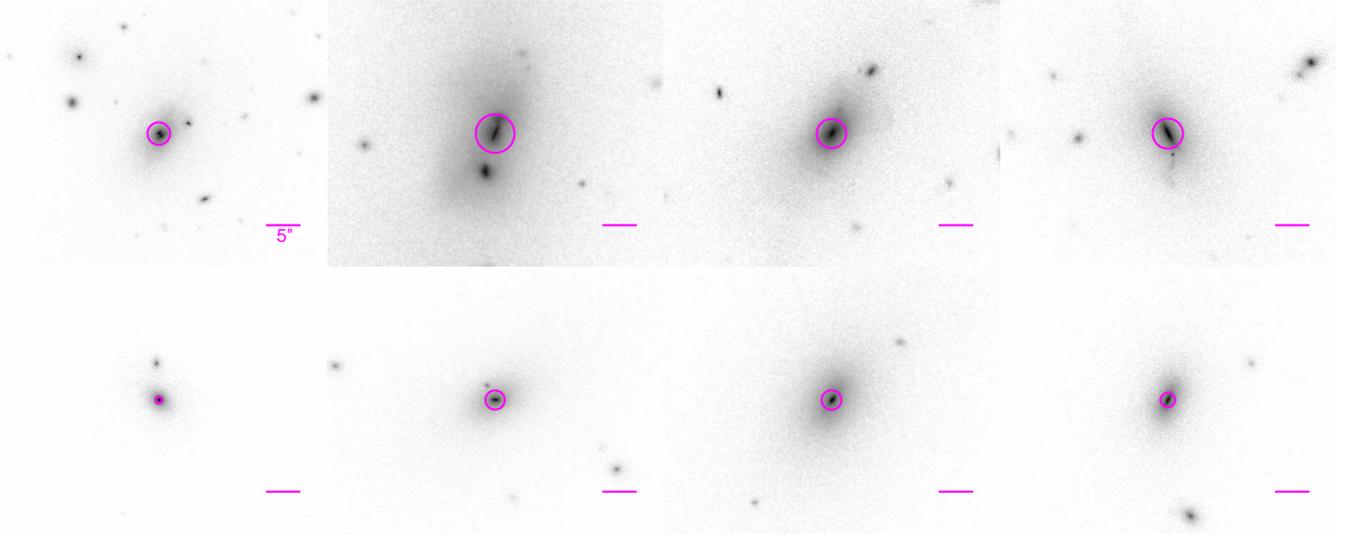

\centering
\includegraphics[width=\linewidth]{{{goodbad.gray}.eps}}
\caption{Stellar surface mass density of representative good and bad lenses. Top row, left to right: $\mathcal{B}_1$, $\mathcal{B}_2$, $\mathcal{B}_3$, $\mathcal{B}_4$. Bottom row, left to right: $\mathcal{G}_1$, $\mathcal{G}_2$, $\mathcal{G}_3$, $\mathcal{G}_4$. The magenta circles denote the Einstein radii, and the scale, denoted by the magenta bars, is identical in all panels. The brightness scale is logarithmic so that the satellite galaxies can be seen.}
\label{fig:goodbad}
\end{figure*}

\begin{figure}
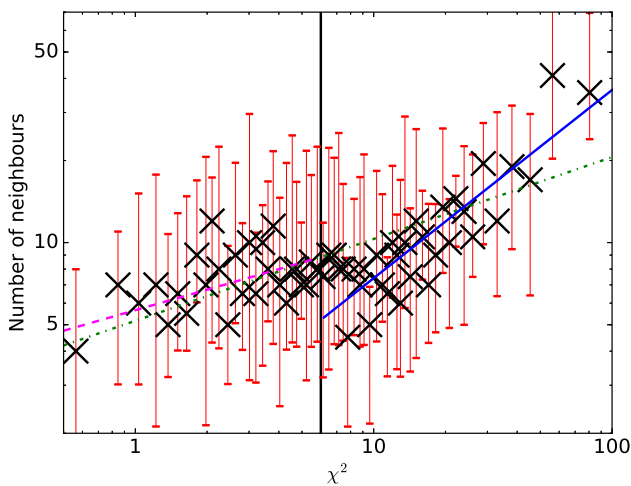

\centering
\includegraphics[width=\linewidth]{{{chi-neigh}.eps}}
\caption{Number of neighbours versus best-fit $\chi^2$ for cusp, fold, and cross lenses (6 degrees of freedom). Each point represents at least 50 lenses, and the error bars represent the intrinsic scatter, containing 68\% of the lenses in that bin. The vertical, black line denotes where the reduced $\chi^2_\nu=1$, and it separates good and bad lenses. The dot-dashed, green line is a power-law fit to all lenses. The dashed, magenta line is fit to the good lenses, and the solid, blue line is fit to the bad lenses.}
\label{fig:chineigh}
\end{figure}

\begin{table}
\centering
\caption{Properties of representative good ($\mathcal{G}_i$) and bad ($\mathcal{B}_i$) lenses. From left to right: Lens, source--lens configuration, reduced $\chi^2$, Einstein radius, minor axis ratio, ratio of $z$-component to total angular momentum, and derived Hubble constant.}
\begin{tabular}{ c c c c c c c }
\hline
lens & conf. & $\chi^2_\nu$ & $\bfrac{r_\textnormal{Ein}}{[\textnormal{arcsec}]}$ & $\frac{c}{a}$ & $\frac{L_z}{|\vec{L}|}$ & $\bfrac{H_0}{[\mathrm{km}\,\mathrm{s}^{-1}\mathrm{Mpc}^{-1}]}$ \\
\hline
\multicolumn{7}{c}{Bad lenses} \\
$\mathcal{B}_1$ & cusp   &  112.2  &  1.59  & 0.23  &  0.97  &  $30.7^{+4.7}_{-3.2}$    \\
$\mathcal{B}_2$ & cross  &  20.7   &  2.75  & 0.51  &  0.74  &  $75.7^{+4.7}_{-5.9}$    \\
$\mathcal{B}_3$ & fold   &  5.7    &  2.09  & 0.64  &  0.04  &  $72.2^{+5.0}_{-4.4}$    \\
$\mathcal{B}_4$ & fold   &  3.1    &  2.15  & 0.49  &  0.98  &  $68.9^{+10.6}_{-9.0}$   \\
\rule{0pt}{0.25ex}   \\
\multicolumn{7}{c}{Good lenses} \\
$\mathcal{G}_1$ & cusp   &  0.5    &  0.50  & 0.71  &  0.21  &  $45.9^{+25.8}_{-17.6}$  \\
$\mathcal{G}_2$ & cross  &  0.4    &  1.37  & 0.72  &  0.52  &  $70.6^{+16.6}_{-13.9}$  \\
$\mathcal{G}_3$ & cross  &  0.2    &  1.40  & 0.72  &  0.98  &  $69.7^{+14.0}_{-14.0}$  \\
$\mathcal{G}_4$ & cross  &  0.2    &  1.03  & 0.60  &  0.98  &  $64.7^{+17.7}_{-14.9}$  \\
\hline
\end{tabular}
\label{tab:goodbadstat}
\end{table}

\begin{table*}
\centering
\caption{Number of lenses and accuracy of model fitting for various configurations at four different lens redshifts ($z_s=1.5$) for two different samples. In each table row 1 gives the lens redshifts. Row 2 gives the configuration of the lens and source. CR, FO, CU, and DO refer to cross, fold, cusp, and double lenses. Row 3 gives the number of lenses in each category. These do not represent unique projections of lens galaxies; e.g. a galaxy may produce lenses in both the cusp and cross categories. The remaining rows give the fraction of lenses for which the corresponding model parameter (see \tabref{tab:modelparams1})is inferred correctly with 68\% confidence. Top: The full sample includes all lenses that pass the selection criteria outlined in the text. Bottom: The good sample includes only those lenses for which the best fit gave a reduced $\chi^2<1$.}
\begin{tabular}{  l || c c c c | c c c c | c c c c | c c c c | c c c c | c c c c | c c c c | c c c c }

\hline
\hline

\multicolumn{17}{c}{Full sample} \\
\hline
$z_\textnormal{lens}$ & \multicolumn{4}{c}{0.183} & \multicolumn{4}{c}{0.366} & \multicolumn{4}{c|}{0.615} & \multicolumn{4}{c}{0.865} \\
configuration & CR & FO & CU & DO & CR & FO & CU & DO & CR & FO & CU & DO & CR & FO & CU & DO \\
\hline
number & 207 & 105 & 21 & 288 & 168 & 138 & 36 & 268 & 103 & 113 & 25 & 211 & 94 & 74 & 19 & 175 \\
\hline
mass & 0.90 & 0.73 & 0.57 & 0.80 & 0.93 & 0.81 & 0.83 & 0.85 & 0.94 & 0.86 & 0.72 & 0.91 & 0.95 & 0.91 & 0.79 & 0.93 \\
$H_0$ & \bf{0.92} & \bf{0.70} & \bf{0.71} & \bf{0.81} & \bf{0.93} & \bf{0.80} & \bf{0.89} & \bf{0.76} & \bf{0.93} & \bf{0.83} & \bf{0.80} & \bf{0.88} & \bf{0.97} & \bf{0.93} & \bf{0.74} & \bf{0.97} \\
q & 0.33 & 0.33 & 0.29 & 0.72 & 0.21 & 0.22 & 0.11 & 0.66 & 0.22 & 0.29 & 0.32 & 0.73 & 0.35 & 0.35 & 0.53 & 0.80 \\
p & 0.29 & 0.36 & 0.24 & 0.54 & 0.15 & 0.22 & 0.17 & 0.44 & 0.12 & 0.30 & 0.40 & 0.49 & 0.26 & 0.27 & 0.47 & 0.64 \\
$\theta_\textnormal{xy}$ & 0.60 & 0.54 & 0.52 & 0.67 & 0.64 & 0.62 & 0.72 & 0.73 & 0.64 & 0.54 & 0.44 & 0.67 & 0.64 & 0.64 & 0.63 & 0.72 \\
$\phi_\textnormal{z}$ & 0.48 & 0.59 & 0.52 & 0.57 & 0.47 & 0.48 & 0.56 & 0.57 & 0.48 & 0.53 & 0.60 & 0.61 & 0.49 & 0.54 & 0.63 & 0.61 \\
x & 1.00 & 0.98 & 0.95 & 1.00 & 1.00 & 0.98 & 1.00 & 1.00 & 0.99 & 0.99 & 1.00 & 1.00 & 0.99 & 0.99 & 1.00 & 1.00 \\
y & 1.00 & 0.98 & 1.00 & 1.00 & 1.00 & 0.99 & 0.97 & 1.00 & 0.96 & 0.99 & 1.00 & 1.00 & 0.99 & 0.99 & 1.00 & 1.00 \\
$r_\textnormal{eff}$ & 1.00 & 0.98 & 1.00 & 1.00 & 1.00 & 0.99 & 1.00 & 1.00 & 1.00 & 1.00 & 1.00 & 1.00 & 0.99 & 1.00 & 1.00 & 1.00 \\

\hline
\hline

\multicolumn{17}{c}{Good sample} \\
\hline
$z_\textnormal{lens}$ & \multicolumn{4}{c}{0.183} & \multicolumn{4}{c}{0.366} & \multicolumn{4}{c|}{0.615} & \multicolumn{4}{c}{0.865} \\
configuration & CR & FO & CU & DO & CR & FO & CU & DO & CR & FO & CU & DO & CR & FO & CU & DO \\
\hline
number & 129 & 17 & 0 & 219 & 132 & 15 & 4 & 224 & 79 & 22 & 0 & 171 & 85 & 34 & 13 & 146 \\
\hline
mass & 0.95 & 0.94 & -- & 0.79 & 0.98 & 0.93 & 1.00 & 0.85 & 0.99 & 0.91 & -- & 0.93 & 0.99 & 0.97 & 0.77 & 0.93 \\
$H_0$ & \bf{0.96} & \bf{0.94} & \bf{--} & \bf{0.81} & \bf{0.98} & \bf{0.93} & \bf{1.00} & \bf{0.76} & \bf{0.99} & \bf{0.91} & \bf{--} & \bf{0.91} & \bf{1.00} & \bf{0.97} & \bf{0.77} & \bf{0.98} \\
q & 0.26 & 0.35 & -- & 0.69 & 0.23 & 0.27 & 0.00 & 0.64 & 0.27 & 0.32 & -- & 0.74 & 0.35 & 0.26 & 0.62 & 0.79 \\
p & 0.21 & 0.41 & -- & 0.52 & 0.16 & 0.33 & 0.50 & 0.46 & 0.13 & 0.41 & -- & 0.48 & 0.26 & 0.26 & 0.62 & 0.62 \\
xyang & 0.62 & 0.82 & -- & 0.68 & 0.65 & 0.80 & 1.00 & 0.74 & 0.68 & 0.45 & -- & 0.68 & 0.66 & 0.68 & 0.69 & 0.73 \\
zang & 0.46 & 0.65 & -- & 0.57 & 0.50 & 0.40 & 1.00 & 0.58 & 0.48 & 0.45 & -- & 0.62 & 0.52 & 0.56 & 0.54 & 0.60 \\
x & 1.00 & 1.00 & -- & 1.00 & 1.00 & 1.00 & 1.00 & 1.00 & 1.00 & 1.00 & -- & 1.00 & 1.00 & 1.00 & 1.00 & 1.00 \\
y & 1.00 & 1.00 & -- & 1.00 & 1.00 & 1.00 & 1.00 & 1.00 & 1.00 & 1.00 & -- & 1.00 & 1.00 & 1.00 & 1.00 & 1.00 \\
$r_\textnormal{eff}$ & 1.00 & 1.00 & -- & 1.00 & 1.00 & 1.00 & 1.00 & 1.00 & 1.00 & 1.00 & -- & 1.00 & 1.00 & 1.00 & 1.00 & 1.00 \\

\hline

\end{tabular}

\label{tab:galstat2}
\end{table*}

\begin{figure*}
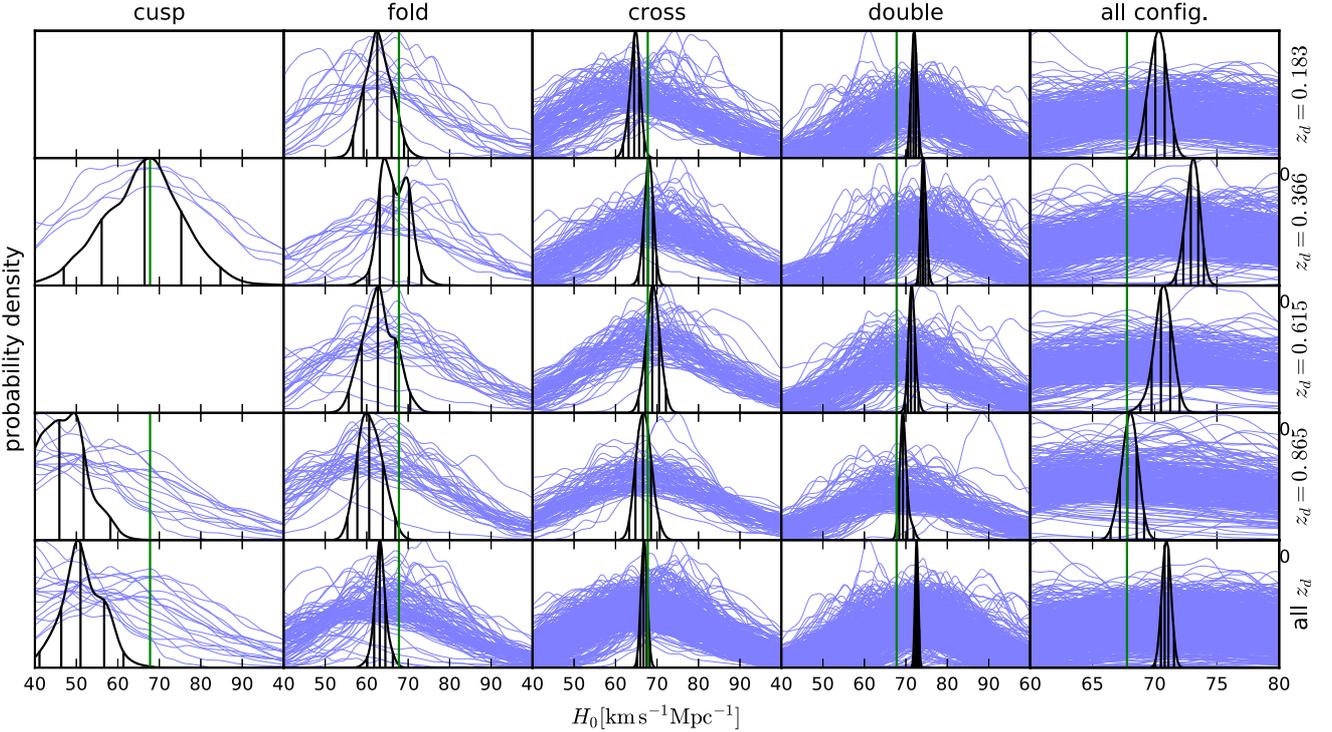

\centering
\includegraphics[width=\linewidth]{{{hist.H0}.eps}}
\caption{Marginalized posterior probability distributions for $H_0$ from lenses in the good sample. Blue curves represent individual lenses, while the black curves are the combined probability distributions. The vertical green lines denote the true value of $H_0$ used in the simulation. The top four rows correspond to different lens redshifts ($z_s=1.5$), while the bottom row includes all lenses from all redshifts. The leftmost four columns correspond to different lens and source configurations, while the rightmost column includes lenses of all configurations. The number of lenses in each redshift--configuration combination are given in \tabref{tab:galstat2}. Note that the curves do not represent unique projections of lens galaxies; e.g. a galaxy may produce lenses in both the cusp and cross categories.}
\label{fig:histH0}
\end{figure*}

\begin{table*}
\centering
\caption{Median values and scatter (symmetrized 68\% confidence interval) of bias on $H_0$, computed from the combined marginalized posterior probability distributions of lenses in Fig.~\ref{fig:histH0}. The number of lenses in each redshift--configuration combination are given in \tabref{tab:galstat2}.}
\begin{tabular}{ c c c c c c }
\hline
$z_d$ & cusp & fold & cross & double & all \\
\hline
0.183 & $0.000\pm0.000$ & $0.923\pm0.050$ & $0.952\pm0.019$ & $1.060\pm0.010$ & $1.034\pm0.011$ \\
0.366 & $0.981\pm0.142$ & $0.981\pm0.052$ & $1.000\pm0.017$ & $1.093\pm0.009$ & $1.076\pm0.009$ \\
0.615 & $0.000\pm0.000$ & $0.925\pm0.060$ & $1.016\pm0.024$ & $1.051\pm0.012$ & $1.040\pm0.011$ \\
0.865 & $0.677\pm0.091$ & $0.894\pm0.045$ & $0.983\pm0.028$ & $1.023\pm0.014$ & $1.000\pm0.010$ \\
all   & $0.753\pm0.077$ & $0.932\pm0.020$ & $0.985\pm0.011$ & $1.069\pm0.007$ & $1.045\pm0.004$ \\
\hline
\end{tabular}
\label{tab:H0stats}
\end{table*}

\begin{figure}
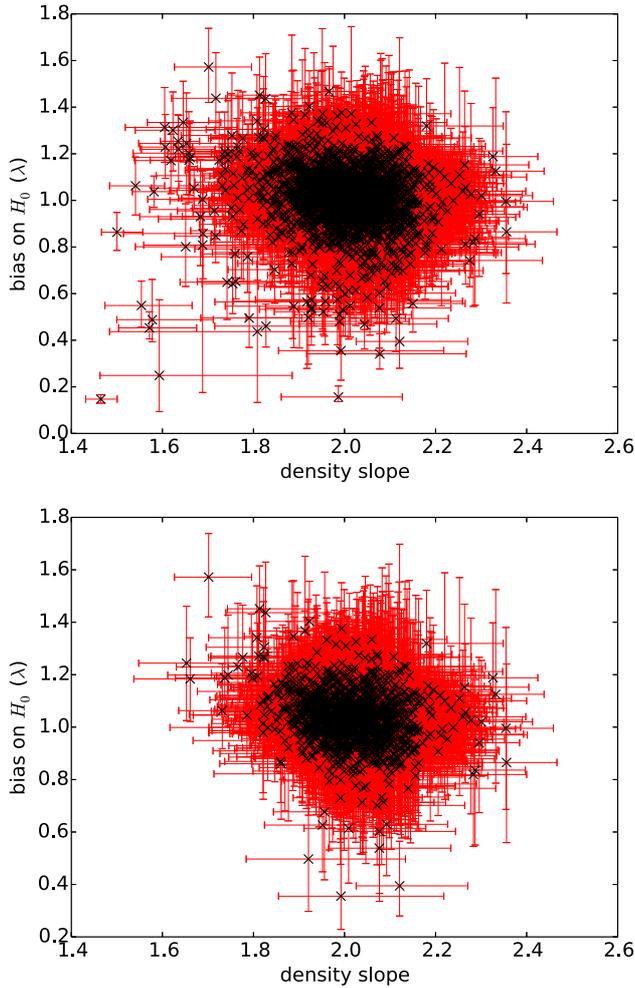

\centering
\begin{tabular}{c}
\includegraphics[width=\linewidth]{{{s-lambda}.eps}} \\
\includegraphics[width=\linewidth]{{{s-lambda.cut}.eps}}
\end{tabular}
\caption{Bias on $H_0$ (ratio of the estimated Hubble constant to true value used in simulation) versus three-dimensional density slope, $\gamma$, with median values and 68\% confidence intervals. The lensing and kinematic data were jointly modelled. Neither the quasar host galaxy nor the lens environment are accounted for. The upper (lower) panel includes lenses from the full (good) sample set, as discussed in the text.}
\label{fig:slambda}
\end{figure}

As previously mentioned, we use the lensed image positions and time delays, along with the aperture velocity dispersion, to constrain the model parameters.
We also include several other observational constraints; the complete list along with assumed observational uncertainties is given in \tabref{tab:modelparams2}.
Because the lensing and kinematic data are assumed to be measured independently from one another, the joint likelihood is the product of the individual likelihoods.
We can therefore write the total $\chi^2$ for a given lens for a given set of model parameters as 
\begin{equation}
\chi^2 = \chi^2_{\textnormal {pos}} + \chi^2_{\textnormal {tdel}} + \chi^2_{\textnormal {dyn}} + \chi^2_{\textnormal {prop}},
\end{equation}
where the terms on the right-hand side are the individual contributions to the total $\chi^2$ due to the (from left to right) image positions, time delays, kinematic constraints, and observable lens properties ($g_x$, $g_y$, $r_\textnormal{eff}$, $e$, and PA).
When all observables and model parameters are considered, there are six (one) degrees of freedom for quad (double) lenses.

In order to compare to the work of \xuabbr and the similar analysis of \textsc{eagle} lenses, we categorize the lenses into two sample sets: the full sample and the good sample.
The full sample contains all lenses, while the good sample contains only those lenses for which a reduced $\chi^2$ (denoted $\chi^2_\nu$) fit of one or less was found for the best set of model parameters.
For simplicity, we refer to lenses from the good sample as good lenses and lensed only found in the full sample as bad lenses.
The sets are further categorized by the lens--source configuration: cusp, fold, cross, or double \citep[see e.g.][for a discussion of lens morphologies.]{saas}.
We find no strong correlation of the $\chi^2$ fit with any lens property (either ``observed'' or extracted from the simulation).
However, Fig.~\ref{fig:chineigh} shows, especially for the bad lenses, a strong correlation of the goodness-of-fit with lens environment.
We performed a least-squares, power-law fit, given by
\begin{equation}
\ln(N_\textnormal{neigh})=m\ln(\chi^2_\nu)+b,
\end{equation}
where $N_\textnormal{neigh}$ is the number of neighbours.
Because of the irregularly-spaced bins, we also weight the data points by the inverse of the local density of points.
The fits give $m=0.30\pm0.03$ and $b=2.18\pm 0.04$ for all lenses, $m=0.25\pm0.06$ and $b=2.18\pm 0.09$ for lenses with $\chi^2_\nu<1$, and $m=0.69\pm0.07$ and $b=1.65\pm 0.09$ for lenses with $\chi^2_\nu>1$.
Not surprisingly, the lens environment plays an important role in the ability of the model to fit the data, and we explore its effects further in section~\ref{sec:adv}.

Fig.~\ref{fig:goodbad} shows several examples of good and bad lenses, and \tabref{tab:goodbadstat} lists some of their properties.
Because double lenses can more easily be fit than quad lenses, these representative lenses were selected based on the $\chi^2_\nu$ fit to either cross, cusp, or fold configurations.
There are many possible reasons why the model could have provided a poor fit to the data; one likely reason is the presence of a satellite galaxy near the lens that needs to be accounted for explicitly in the lens model.
$\mathcal{B}_1$ appears to be a merging system, while $\mathcal{B}_2$ has a massive companion within the multiply-imaged region.
There is a less massive companion near the Einstein radius of $\mathcal{B}_3$, but there is also a massive object at a distance of $\sim 5r_\textnormal{Ein}$.
$\mathcal{B}_4$ appears to have undergone a recent minor merger, as evidenced by a stellar stream.
It also has a flattened, disk-like morphology and was classified as a fast rotator.
The good lenses are typically in less crowded environments and have fewer companions near the Einstein radius.
For this reason, they also have systematically smaller Einstein radii.
Nevertheless, there are still some systems, such as $\mathcal{G}_2$, that have satellites inside the multiply-imaged region but are still well-fit by the model.

In \tabref{tab:galstat2} for each sample set, we quote the total number of lenses and attempt to quantify the fidelity of the model fitting.
As the lenses are drawn from the simulation, we know the true values of several key parameters that are directly fit, such as the shape and orientation of the galaxy, and we list the fraction of galaxies for which the true values are recovered at 68\% confidence.
Because of the small number of cusp and fold lenses, it is difficult to make a direct comparison between the full and good sample sets. 
However, in general we note a consistent increase in the fraction of good lenses for which the mass and $H_0$ are recovered.
The axis ratios and orientation of the galaxy, on the other hand, show poor fits across many categories. 
This discrepancy is most likely due to a combination of inaccurate estimation of these parameters from the raw particle data and inaccurate modelling of the lens environment.
In the simulations, the size of the spherical aperture in which these shape parameters are measured will significantly affect the fit.
The presence of a central bulge, disk, or a nearby/merging galaxy will also play a significant role.

Focusing on the good sample set, we separate the lenses by redshift and configuration. 
Fig.~\ref{fig:histH0} shows the marginalized posterior probability distribution for $H_0$, for each lens, and \tabref{tab:H0stats} shows the medians and scatter for each redshift--configuration combination.
Combining results across all redshifts, the cross lenses show the least bias among all lenses.
The fold lenses are biased low, and the double lenses are biased high.
It is more difficult to come to a conclusion about the cusp lenses.
The good sample contains no cusp lenses at two redshifts: $z_d=0.183$ and $z_d=0.615$. 
Cusp lenses at $z_d=0.366$ show no bias and peak near the correct value of $H_0$.
On the other hand, there are $\sim 10$ cusp lenses at $z_d=0.865$, which all peak at low values of $H_0$.
There are, unfortunately, too few cusp lenses to generalize these results.
The method used to produce lenses from the simulation was to place quasars uniformly behind lenses (see \appref{app:pdp2}), creating mock observations for every quasar.
Because the region of the source plane that can produce cusp lenses is the smallest of all quad configurations, the method was not optimal for producing a large number of cusp lenses.
Nevertheless, cusp lenses are morphologically similar to double lenses.
Three of the images in a cusp configuration are very near one another and, accordingly, have similar time delays.
Thus, one can expect them to behave similarly to double lenses.

Next we compare the analysis of section~\ref{sec:comparison} to this analysis.
Fig.~\ref{fig:slambda} shows the scatter in the $s$--$H_0$ plane.
Using the full sample set, there is a slight correlation between density slope and the bias on $H_0$, suggesting that isothermal fits perform better, but focusing on the good sample set shows that the shallower density profiles associated with lower values of $H_0$ are not well-described by a power-law alone.
The strong correlation in the $s$--$H_0$ plane seen in section~\ref{sec:comparison} is not present here; the remaining lenses that are biased do not show any preference for density slope.

As we have seen, the $\chi^2$ fit depends on the lens environment.
The good sample, defined as those lenses with a reduced $\chi^2<1$, thus naturally corresponds to less crowded environments.
As a sanity check, we have also confirmed that, when compared to the full sample, the recovered model parameters from the good sample are more consistent with those corresponding parameters extracted from the simulation.
Focusing on the good sample, the double lenses are biased at the 5\% level with an intrinsic scatter of 10\%, which is similar to the case without kinematics.
The quad lenses, on the other hand, are biased only at the 0.5\% level with a 10\% intrinsic scatter, which represents a significant reduction in bias.
As suggested by Fig.~\ref{fig:chineigh} and section~\ref{sec:adv}, taking into account the lens environment may play a significant role in further reducing the bias and scatter in $H_0$ estimates.

\subsubsection{Effect of velocity dispersion constraints}
\label{sec:nokin}

As seen in section~\ref{sec:results2}, the use of lensing observables and kinematic data improves estimates of $H_0$ significantly.
Compared to the results of \xuabbrns, the bias itself is driven towards unity, and the scatter is minimized without having to perform any additional selection cuts on the good sample.
However, the lens samples in section~\ref{sec:results1} and section~\ref{sec:results2} are not identical.
Moreover, the former section assesses the bias by fitting a power-law to the convergence after an MST is applied; the latter fits the observables and infers $H_0$ directly.
Because of these significant differences, it is not clear what impact the inclusion of kinematic information has had on the inference of $H_0$.

To address this question, we focus only on the double lenses since they are also analogous to the lenses of section~\ref{sec:comparison}, where it is assumed that only two lensed images are observed.
However, this time we ignore any kinematic information.
Because the double lenses are already biased towards high values of $H_0$ with the inclusion of kinematic data, we do not necessarily expect these estimates to improve on average without the constraints.
We do, however, expect the uncertainty on the derived Hubble constant to increase.

We combine the posterior probability distributions from double lenses across all redshifts.
With no kinematic information, we find that bias on $H_0$ (i.e., the ratio of the measured Hubble constant to the true value used in the simulation, $H_{0,\textnormal{measured}}/H_{0,\textnormal{true}}$) is 1.080 and the 68\% and 95\% confidence intervals are [1.071,1.089] and [1.063,1.098], respectively.
After adding kinematic constraints, the recovered bias is $H_0=1.055$, and the confidence intervals are [1.050,1.060] and [1.046,1.065].
There is still a significant bias in both cases, but the bias has slightly reduced with the addition of velocity dispersion information.
Also, the uncertainties have nearly halved after the inclusion of kinematic data, since there are now more observational constraints.

\section{Lens environment and host galaxy}
\label{sec:adv}

\begin{table}
\centering
\begin{tabular}{l | c c c}
Lens & Time delays & +group & +group+host \\
\hline
$\mathscr{L}_1$ & $1.366^{+0.435}_{-0.388}$ & $1.039^{+0.360}_{-0.322}$ & $1.027^{+0.063}_{-0.053}$  \\
$\mathscr{L}_2$ & $1.164^{+0.245}_{-0.218}$ & $0.980^{+0.217}_{-0.217}$ & $0.981^{+0.109}_{-0.124}$ \\
\hline
\end{tabular}
\caption{Median values for bias on Hubble constant and confidence intervals for $\mathscr{L}_1$ and $\mathscr{L}_2$. The bias is given by the measured value of $H_0$, scaled by the true value used in the simulation.}
\label{tab:L1L2}
\end{table}

\begin{figure*}
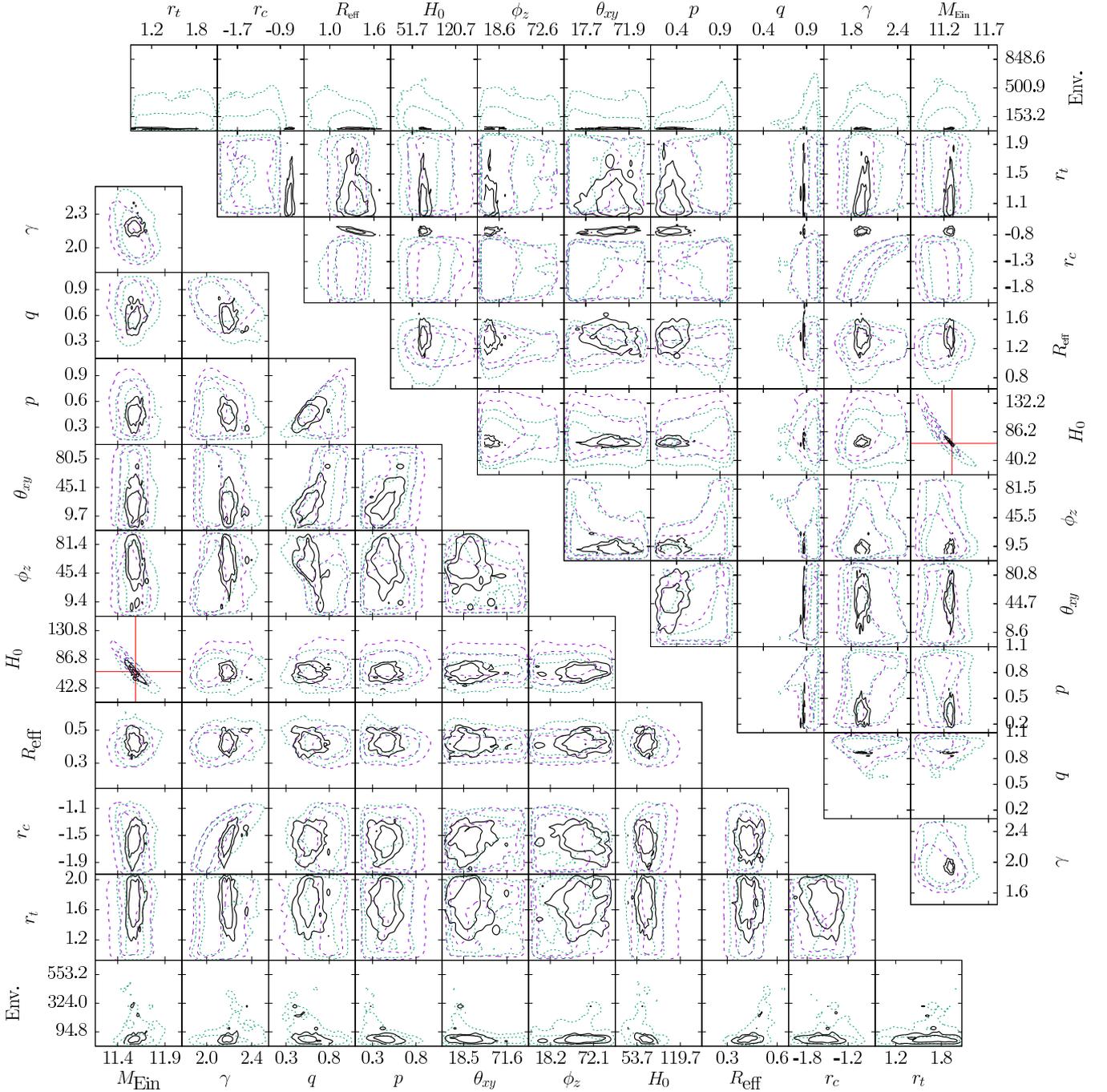

\centering
\includegraphics[width=\linewidth]{{{cov}.eps}}
\caption{Two-dimensional joint probability distribution for all model parameters. The upper (lower) triangular plots correspond to $\mathscr{L}_1$ ($\mathscr{L}_2$). The long-dashed, magenta contours take into account the quasar and lens galaxy observables. The short-dashed, green contours result from also including the lens environment. The solid, black lines include constraints from the lens environment and quasar host galaxy as well. The horizontal and vertical red lines denote the true values of $H_0$ (used in the simulation) and the mass inside $R_\textnormal{Ein}$ (extracted from the particle data). See \tabref{tab:modelparams1} for a description of the model parameters. The Env. parameter controls the overall normalization of the neighbouring galaxies; it is the Einstein radius, in mas, of the most massive neighbour.}
\label{fig:extsbcorner}
\end{figure*}

The previous analyses ignored some information that would be readily available from observations.
Lens galaxies are rarely isolated, and nearby galaxies can have a significant effect on the observables.
Furthermore, mass along the line-of-sight must be considered as well.
If there is an external sheet of convergence $\kappa_\textnormal{ext}$ that is unaccounted for, the Hubble constant will be biased high by a factor of $(1-\kappa_\textnormal{ext})^{-1}$.
For example, the double lenses are biased high, which could be explained if there is an unaccounted group halo or a significant number of neighbours nearby.
Here, we take into account massive satellite galaxies near the lens, as well as the extended surface brightness distribution of the lensed host galaxy, which we presume is observable.
We do not consider line-of-sight mass because the \textsc{eagle} simulation box is not large enough to generate such effects.

To account for the lens environment, we adopt a simplistic approach in which all visible nearby galaxies are modelled as singular isothermal spheres, characterized by a constant velocity dispersion $\sigma_\textnormal{SIS}$.
However, to keep the number of parameters reasonable, we assume these galaxies are all early-types and follow a Faber-Jackson relation.
We assume a constant mass-to-light ratio for each of these galaxies so that $\sigma_\textnormal{SIS}^4 \propto L_M$, where $L_M$ is the luminosity, and mass is used as a proxy for luminosity.
We identify nearby galaxies and estimate masses using \textsc{sextractor}.\footnote{We use a minimum detection threshold of 4\e{9} $M_{\astrosun}/\textnormal{mas}^2$ and a minimum area of 12.5 square mas, which were chosen so that, by visual verification, most galaxies were uniquely selected by \textsc{sextractor} as a single object.}
The algorithm detects objects, which, by visual inspection, are not truly galaxies. 
However, they are typically detected with relatively low stellar masses and will therefore not significantly affect the lensing observables.
Because the Faber-Jackson relation now fixes the velocity dispersions of the satellite galaxies fixed with respect to one another, there is only one free parameter, which we take as the velocity dispersion of the brightest satellite.

Another key piece of information is the extended light distribution of the quasar host galaxy.
Although reconstructing the host galaxy does not provide direct constraints on $H_0$, it can tightly constrain the parameter space in other dimensions and remove degeneracies.
For each quasar, we create mock data of the lensed host by placing a galaxy with an elliptical S\'{e}rsic light profile, with random S\'{e}rsic index, shape and orientation,\footnote{The S\'{e}rsic index is limited to the range 0.5--1.5. The ellipticity is limited to 0--0.5, and the scale radius is fixed to 0.1 arcsec.} at the position of the quasar.
The data are constructed on a grid with a pixel scale of 0.05 arcsec/pixel, convolved with a circular Gaussian point spread function of standard deviation one pixel, and Gaussian noise is added so that the peak signal-to-noise is 25. 
We apply the pixel-based and shapelets-based source reconstruction methods described in \citet{pbsr1} and \citet{pbsr2}, respectively; we find no significant difference in using either technique.
Below, we use $15\times 15$ shapelets and curvature regularization (optimized within a Bayesian framework) to derive lens model parameter uncertainties, image-plane model residuals, and source reconstructions.

A complete analysis of all the \textsc{eagle} lenses in such a manner is beyond the scope of this current work.
However, to demonstrate the possibility of reducing the bias on $H_0$, we analyze two randomly-chosen double lenses, denoted $\mathscr{L}_1$ and $\mathscr{L}_2$, which meet several criteria.
They are not classified as fast rotating galaxies (see section~\ref{sec:lenssample2}), are not merging or recently merged, have at least five neighbouring galaxies, have an Einstein radius of at least 1 arcsec, and show a bias on $H_0$ of $\gtrsim 1\sigma$.
For reference, the environment of the lenses out to $R_{200}$ and the input data are shown in Fig.~\ref{fig:extsb}.

Because of the lensing equation can be written in a dimensionless form, there is little to no sensitivity of the Hubble constant to the quasar host galaxy reconstruction. 
Nevertheless, the large number of observational constraints (i.e., the number of pixels in the annulus of the strong lensing region of interest) available when modelling the host galaxy allows certain model parameters, such as the density slope, to be tightly constrained.
Therefore, including the host galaxy in the analysis can still break degeneracies and tighten constraints on $H_0$.
We show two-dimensional joint posterior probability distributions for the model parameters in Fig.~\ref{fig:extsbcorner}, separately highlighting those from the time delays alone; the time delays and the group environment; and the time delays, group, and host galaxy.
\tabref{tab:L1L2} gives the estimates of the Hubble constant for these different analyses.

We chose lenses for which $H_0$ is overestimated when not considering the group environment.
After factoring the nearby galaxies into the lens model, this bias has disappeared for both lenses, and the uncertainties have shrunk by $\sim$15--20\%.
After combining constraints from the quasar host galaxy, the uncertainties shrink even further (relative to the case with no group or host galaxy constraints) by factors of approximately 7 and 2 for $\mathscr{L}_1$ and $\mathscr{L}_2$, respectively.
The drastic improvement seen for $\mathscr{L}_1$ is largely due to the presence of a central image of the host galaxy.
This allows the core radius and density slope to be tightly constrained.
The uncertainties on the truncation radius are still large, but the mass inside the Einstein radius is relatively insensitive to $r_t$ in this case.
Consequently, the large degeneracy in the $M_\textnormal{Ein}$--$H_0$ plane is significantly reduced.

\section{Conclusions}
\label{sec:conclusion}

Using galaxies in the \textsc{eagle} simulation, we have investigated the bias on the Hubble constant estimated using strong lensing.
In section~\ref{sec:comparison}, we perform an analysis similar to that of \xuabbr in which a mass-sheet transformation (MST) is applied to the radial density profile of lens galaxies, and we arrive at similar key results.
Lenses with larger (SIS) velocity dispersions tend to show the least bias and scatter (Fig.~\ref{fig:fig8}).
There is also a strong correlation between density slope and bias, but by selecting those lenses that are nearly isothermal, the bias and scatter can be significantly reduced (Fig.~\ref{fig:fig9}; \tabref{tab:ssl}).
On the other hand, one of the key differences between the \textsc{eagle} galaxies and the Illustris galaxies used by \xuabbr lies in their sizes. 
The \textsc{eagle} galaxies are calibrated to reproduce the observed galaxy mass--size relationship and are on average smaller in size than the Illustris lenses, since Illustris overestimates galaxy sizes.
Consequently, \textsc{eagle} lenses have, on average, smaller mean, median, and scatter in their Einstein radii (\tabref{tab:galstat}).
The equivalence radii (the radii at which dark matter begins to dominate over baryons) is also smaller for \textsc{eagle} lenses; almost all equivalence radii we measure are within their corresponding Einstein radii (Fig.~\ref{fig:fig7}).

We next investigated whether combining lensing observables with kinematic constraints can further reduce uncertainties by using the \textsc{eagle} simulation to create mock observations of lensed quasars and to extract central velocity dispersions. 
By combining the two constraints and focusing on those lenses which are well-described by a cored and truncated power-law model, we are able to significantly reduce the bias and scatter on $H_0$.
The correlation of the bias with density slope also disappears (Fig.~\ref{fig:slambda}).
Classifying the lens configurations as either cross, fold, cusp, or double, we find that cross lenses show the least bias of all lenses (\tabref{tab:H0stats}).
Fold lenses seem to be biased low, while double lenses are biased high (Fig.~\ref{fig:histH0}).
For double lenses, the bias and intrinsic scatter are 6\% and 10\%, respectively, while for quad lenses, the bias and intrinsic scatter are 0.5\% and 10\%, respectively.
We note that the environment of the lens may play a significant role in improving measurements of $H_0$ (Fig.~\ref{fig:chineigh}).

Focusing on two double lenses which show significant bias and satisfy several criteria, we attempt to model the data in more detail.
We take into account the effect of the lens environment on the modelling.
By identifying massive objects near the lens galaxy and using a simple prescription to relate the measured masses to SIS models for each object, we directly model the galaxies along the line of sight and find that the bias significantly decreases for these two lenses.
We also simulate the extended emission from the quasar host galaxy and model this mock data using source reconstruction techniques.
The addition of the host galaxy does not introduce any bias and significantly reduces the statistical uncertainty on $H_0$ (\tabref{tab:L1L2}).
Modelling the extended emission cannot itself constrain $H_0$.
However, when combined with time delay measurements, as we have done, modelling the extended surface brightness can break degeneracies in the multi-dimensional parameter space and tighten constraints on $H_0$ (Fig.~\ref{fig:extsbcorner}).

Strong lensing as a tool to probe cosmology is already proving to be complementary and competitive.
In the current work, we find that cross lenses are the least biased of all source--lens configurations.
Except for the lowest lens redshift, the true Hubble constant was recovered at all redshifts using cross lenses.
Across all redshifts using 425 lenses, an unbiased 1\% precision was achieved for cross lenses.
As evidenced from two particular double lenses, careful modelling of the host galaxy and environment can remove bias and reduce statistical uncertainties. 
This suggests that by focusing on cross lenses and carefully modelling the host and environment, as well as accounting for additional systematics such as mass along the line-of-sight, cosmological constraints from strong lensing can be competitive. 

\section{Acknowledgements}
\label{sec:ack}

This work used the DiRAC Data Centric system at Durham University, operated by the Institute for Computational Cosmology on behalf of the STFC DiRAC HPC Facility (www.dirac.ac.uk). This equipment was funded by BIS National E-infrastructure capital grant ST/K00042X/1, STFC capital grants ST/H008519/1 and ST/K00087X/1, STFC DiRAC Operations grant ST/K003267/1 and Durham University. DiRAC is part of the National E-Infrastructure.
This work was  supported by the Netherlands Organisation for Scientific Research (NWO), through VICI grant 639.043.409 and by the European Research Council under the European Union's Seventh Framework Programme (FP7/2007- 2013) / ERC Grant agreement 278594-GasAroundGalaxies.
We also note that this work used support from the UK Science and Technology Facilities Council in the form of a postdoctoral research assistantship.

\bibliographystyle{mnras}
\bibliography{refs}

\appendix

\section{Raw data processing and methods}

\subsection{Radial density calculations}
\label{app:pdp1}

There are many possible definitions that can be used to define the edge of a lens galaxy.
When identifying a galaxy or the group within which a galaxy resides, we select all particles inside $R_{200}$, which is the spherical radius within which the mean density is 200 times the critical density of the Universe.
For each projection along a coordinate axis, the discrete particle positions must be converted into a smooth representation of the data.
The exact procedure depends on the nature of the simulation.

SPH simulations describe the motions and interactions of fluids using a finite number of discrete particles.
When projecting the particles to make maps, the properties (e.g. mass, velocity) are convolved with a smoothing kernel, and each particle has its own smoothing length.
When map making we use a cubic spline kernel with finite extent for all particles.
The smoothing length is defined as the radius of the sphere, centered on a given particle, which encompasses 42 other particles.
Thus, particles in more (less) dense regions have smaller (larger) smoothing lengths and occupy a smaller (larger) volume of the fluid.

After determining the smoothing lengths, all particles along a given viewing axis are projected onto a two-dimensional plane. 
The plane is divided into pixels, and the overlap of each particle's projected smoothing kernel with the grid is computed. 
The quantity of interest can then be weighted appropriately for each particle and pixel.

When evaluating the convergence in section~\ref{sec:comparison}, we smooth all particles within $R_{200}$ onto a grid with pixel size of 50 physical pc.
The width of the grid is the larger of either $6r_{\textnormal{eff},0}$ or $7r_{\textnormal{Ein},0}$.
Here, $r_{\textnormal{eff},0}$ is the effective radius, which encloses half the projected mass, and it is estimated (before smoothing and projection) from those particles within 30 kpc.
$r_{\textnormal{Ein},0}$ is the circularized Einstein radius, estimated from the particle data, and is the radius which satisfies
\begin{equation}
\frac{M(<r_{\textnormal{Ein},0})}{\pi \Sigma_\textnormal{cr}r_{\textnormal{Ein},0}}=r_{\textnormal{Ein},0}, 
\end{equation}
where $M(<r_{\textnormal{Ein},0})$ is the mass within $r_{\textnormal{Ein},0}$.
We note that because we examine various lens/source redshift combinations, $\Sigma_\textnormal{cr}$, the critical surface mass density for lensing, itself varies with redshift.

The width and pixel scale of the grid ensure the convergence map is well-resolved in the relevant regions, down to the softening length.
Thus, to extract the radial profile of the lenses, a tenth-degree polynomial is fit (in log--log space) to the convergence map, with appropriate weighting to account for the larger number of pixels at larger radii, from the softening length to the maximum radius. 
Evaluation of the density below this region is only necessary when computing the cumulative density distribution $\bar{\kappa}$.
For this, we perform a linear fit in log-space to the radial profile between 1--1.5 times the softening length and extrapolate to smaller radii.

\subsection{Extracting lensing and kinematic observables from the simulation}
\label{app:pdp2}

For the analysis described in section~\ref{sec:dataextract}, our general approach for smoothing and projecting particles onto a grid is identical to that described in \appref{app:pdp1}.
The choice of grid size and pixel scale is also the same unless otherwise noted.

For calculating the half light radius $r_{\textnormal{eff}}$, we use an iterative approach. 
Without visual inspection, it can be difficult to know whether there exist any massive objects near the galaxy of interest. 
A first estimate of $r_{\textnormal{eff}}$ is made from the stellar particle data.
Star particles within a spherical aperture of 30 kpc are projected (but not smoothed), and $r_{\textnormal{eff}}$ is given by the projected radius that encloses half of the mass.
This gives an estimate of the effective radius, but the 30 kpc aperture could exclude a significant fraction of particles for larger galaxies.
On the other hand, it could be too large for smaller galaxies and include neighbouring companions.
We therefore project and smooth the stellar particles out to five times our original estimate.
$r_{\textnormal{eff}}$ is then determined from the smoothed map, by considering all pixels within a radius that is at least 20 kpc and at most four times the original estimate.
This procedure will still fail for some galaxies with a nearby neighbouring galaxy, but we expect it to perform better than the original estimate.

We calculate the lensing potential and deflections on square grids that are $6 r_\textnormal{Ein}$ on each side and have pixel scales of 10 mas; $r_\textnormal{Ein}$ is calculated as before.
Although the lens redshift varies, the source redshift is fixed at $z_s=1.5$.
The potential and deflections can be related to the convergence by
\begin{equation}
\begin{aligned}
\phi(\mathbf{n}) = &\frac{1}{\pi}\sum_\mathbf{m}\log\lvert \mathrm{D}[\mathbf{n},\mathbf{m}]\rvert\;\kappa(\mathbf{m}) \\
&\textnormal{and} \\
\alpha(\mathbf{n}) = &\frac{1}{\pi}\sum_\mathbf{m}\frac{\mathrm{D}[\mathbf{n},\mathbf{m}]}{\lvert \mathrm{D}[\mathbf{n},\mathbf{m}]\rvert^2}\;\kappa(\mathbf{m}),
\end{aligned}
\label{eq:potdef}
\end{equation}
where the sums are taken over all pixels in the convergence map, $\mathbf{n}$ represents a pixel in the potential or deflection map, and $\mathrm{D}[\mathbf{n},\mathbf{m}]$ is the position vector pointing from $\mathbf{n}$ to $\mathbf{m}$.
For reference we note that the lens equation, Jacobian $A$, and magnification tensor $\mu$ are given by  
\begin{equation}
\begin{aligned}
\vec{u} &= \vec{x}-\vec{\alpha}(\vec{x}), \\
A &= \partial\vec{u}/\partial\vec{x}, \\
&\textnormal{and} \\
\mu &= A^{-1},
\end{aligned}
\end{equation}
respectively. $\vec{u}$ denotes positions in the source plane, while $\vec{x}$ denotes positions in the lens plane.

The smoothed convergence map used in \eqref{eq:potdef} is created from all particles within $R_{200}$.
However, to limit the time needed to compute the potential and deflection, the resolution of the convergence map varies with distance. 
Within the innermost $6r_\textnormal{Ein}\times 6r_\textnormal{Ein}$ region, the pixel scale is 10 mas. 
Out to $12r_\textnormal{Ein}\times 12r_\textnormal{Ein}$ the scale is 20 mas.
The resolution in the remaining region of the map differs from lens to lens but is typically $\sim 200 \textnormal{mas}$.

To generate quad lenses, we examine the lensing Jacobian and identify the smallest rectangle in the source plane that encloses the tangential caustic.
The tangential caustic is identified by moving outwards from the center of the lens and noting where the determinant of the Jacobian matrix changes sign.
To produce a range of lens morphologies while retaining computational efficiency, we then take 2601 uniform samples in the region containing the caustic as source positions of a potentially lensed quasar.
We solve the lens equation and accept the source position as viable if we can identify five images, one of which is demagnified and nearest to the lens center.
More specifically, we minimize the quantity $ L\equiv|\vec{x}-\vec{\alpha}(\vec{x})-\vec{u}|$ for each pixel and identify each minimum as an image position.
Because of the finite mass resolution of the simulation, $L$ will not be exactly zero at the minima; we require that $L<0.05r_\textnormal{Ein}$ for a minimum to be considered a true image position.

We next classify the lens morphologies as being either cusp, fold, cross, or neither; the image separation ratios between all pairs are used to discriminate between the categories.
For cross lenses for all four images, we require the difference of the separations between the two neighbouring images of opposite parity to be less than 20\%.
For fold lenses, we first compute the minimum separation between any pair of images. We then require two of the images (those straddling the critical curve) to be minimally separated from one another and separated by more than twice the minimum separation from the remaining two images. We also require the remaining two images (those not straddling the critical curve) to be separated by more than twice the minimum separation from all other images.
For cusp lenses, we again first compute the minimum separation. We require three of the images (those straddling the critical curve) to be separated from two others by at most 125\% of the minimum separation and separated from the remaining image by more than twice the minimum separation. 
The lensed sources that pass these tests are then inspected by eye to verify their morphologies, and we keep one lens from each category for analysis using the methodology described in section~\ref{sec:modelling}.

\section{Dynamical modelling}
\label{app:mi}

\begin{table*}
\centering
\caption{Standard deviations of interpolation errors for various minor axis ratio bins.}
\begin{tabular}{  l || c c c c c c c c}
\hline
Minor axis ratio ($c/a$) & 0.2--0.3 & 0.3--0.4 & 0.4--0.5 & 0.5--0.6 & 0.6--0.7 & 0.7--0.8 & 0.8--0.9 & 0.9--1.0 \\
\hline
Standard deviation of fractional error & 0.11 & 0.08 & 0.06 & 0.04 & 0.03 & 0.03 & 0.02 & 0.02 \\
Standard deviation of absolute error ($\mathrm{km}\,\mathrm{s}^{-1}$) & 11.74 & 8.07 & 6.51 & 4.69 & 3.73 & 3.68 & 2.89 & 1.88 \\
\hline
\end{tabular}
\label{tab:apptab}
\end{table*}

\subsection{Orbit modelling}

Section~\ref{sec:kinematics} describes the dynamical modelling of lens galaxies.
As described, \textsc{smile} simulates the orbits of particles in the gravitational potential that corresponds to a given mass density.
The resulting N-body snapshot is then used to predict central velocity dispersions.
Before orbit modelling can be done, however, initial particle positions and velocities must be generated.
This is accomplished by first populating the volume with particles that approximate the deprojected S\'{e}rsic profile, which is the density profile we choose to represent the stellar particles. 
Then, initial velocities are derived analytically by solving the Jeans equation for the axisymmetrized version of the true density model (\textsc{smile} does this automatically.).
After the orbits are integrated in time, they are weighted (with a uniform prior and a variable regularization on the weights) so that the desired stellar population is reproduced.
The particle positions, velocities, and weights are then recorded.

The N-body snapshots can be used to calculate line of sight velocity dispersions within an aperture, for any viewing angle.
$10^5$ particles are sufficient for calculating aperture velocity dispersions; typical uncertainties, which are significantly smaller than observational uncertainties, are less than one percent.
Although the particles do not come from an SPH simulation, we find no significant difference between calculating the dispersion from the particle data (root mean square velocity) or from the result of applying a smoothing kernel to the particles (as is done for the \textsc{eagle} galaxies).
Lastly we note that converting from N-body velocity units to physical units requires a rescaling by $\sqrt{G M_\textnormal{inf}/r_{3d}}$, where $G$ is the gravitational constant.
Given a particular choice of model parameters in \tabref{tab:modelparams1}, the total mass and three-dimensional radius can be calculated analytically or numerically.

\subsection{Multidimensional interpolation}

As noted in the text, the Schwarzschild orbit modelling code \textsc{smile} is computationally expensive. 
Therefore, when evaluating the velocity dispersions for a set of lens model parameters, we use radial basis functions to interpolate between pre-tabulated values. 
Specifically, for a set of viewing angles ($\theta_{xy}$ and $\phi_z$), we use the so-called thin-plate spline (TPS), given in \citet{tps}, to interpolate over the remaining relevant parameters: $\gamma$, $q$, $p$, $r_c$, and $r_t$.
Interpolation using the TPS provides higher accuracy for smoothly varying functions, which requires fewer velocity dispersion calculations.

In two dimensions, the TPS is the solution to the biharmonic equation, $\nabla^4 U(r)=0$, whose fundamental solution is given by $U(r)=r^2\ln(r)$.
$U(r)$ is a non-localized ($\lim_{r\to\infty} U(r)=\infty$) radial basis function and thus provides a global fit to the velocity dispersions.
This is attractive from a physical perspective because the TPS, as the name suggests, is the surface that fits a three-dimensional surface such that the bending energy (an integral over the second derivatives of the surface) is minimized.
It can be thought of as the shape a thin rubber sheet would take if fixed at set of given points.
Moreover, the TPS is scale-invariant and requires no manual tweaking of parameters.
In higher dimensions, the TPS no longer minimizes the bending energy, but it is still useful for interpolation.
After interpolation using the TPS, we perform a simpler bilinear interpolation over the viewing angles ($\theta_{xy}$ and $\phi_z$).

To test the accuracy of the interpolation method, we compute velocity dispersions (from the tabulated velocity dispersions) for 1000 sets of model parameters, randomly chosen from the allowed parameter space.
We compare these values to the those obtained from simulations using \textsc{smile}.
The fractional errors depend most strongly on the minor axis ratio and are largest when the minor axis ratio is smallest.

Converting the velocity dispersion to physical units requires a rescaling by $\sqrt{G M_\textnormal{inf}/r_\textnormal{3d}}$.
Because the total mass scales (roughly) with axes ratios as $M_\textnormal{inf} \propto 1/(qp)$, the more triaxial objects have larger total mass.
However, these models also tend to  have much smaller velocity dispersions (in N-body units).
Thus, a large fractional error does not necessarily imply a large absolute error.
To investigate this, each of the 1000 model parameters evaluated is scaled to physical units for 100 randomly masses and scale radii.\footnote{The masses and radii are taken from the MCMC samples of all lenses in the full sample.}
In total, this yields $10^5$ absolute errors.
\tabref{tab:apptab} shows the standard deviations of the fractional and absolute errors for different minor axis ratio values.
The two measures of error are well-correlated, but more importantly, the standard deviation of the absolute errors is typically much less than $10\mathrm{km}\,\mathrm{s}^{-1}$, which is the 1$\sigma$ uncertainty assumed on velocity dispersions when modelling the data.

\section{Identification of fast rotators}
\label{app:fastrotators}

To calculate the minor axis ratios ($c/a$), the ratio of the velocity dispersion along the minor axis to the total velocity dispersion ($\sigma_z/\sigma_\textnormal{tot}$), and angular momentum of each galaxy ($\vec{L}$), as presented in section~\ref{sec:lenssample2}, we use a modified version of the shape measurement code distributed with \textsc{smile}.
Briefly, the particles of a galaxy lying within 30 kpc are placed into 20 radial bins.
For each bin, starting with the smallest bin, the moment of inertia tensor, computed from all particles inside the bin, is diagonalized. 
An elliptical radius, as opposed to the initial spherical radius, is then calculated for each particle, using the eigenvalues of the moment of inertia tensor.
Using the elliptical radius now, the moment of inertia tensor is again diagonalized and the elliptical radii are updated.
This processed is repeated until the eigenvalues have converged.
Additionally, this process is iteratively performed from the smallest bin to the largest bin, so that the algorithm is stable, and only the results from the last bin are kept.
Finally, the galaxy is rotated so that it lies along its principal axes.
Velocity dispersions and angular momenta are computed along or about each axis, using appropriate mass weightings.

We note that \citet{cLz}, hereafter A+16, have investigated similar properties for barred galaxies in the \textsc{eagle} simulation. 
The authors compute, for a subset of the galaxies, the distributions of $c/a$ and $\sigma_z/\sigma_\textnormal{tot}$.
From visual inspection, the range of minor axis ratios A+16 find differs from what we find. 
While A+16 find a significant fraction of lenses with $c/a$ between 0.2--0.5, we find significantly fewer.
On the other hand, the distribution of $\sigma_z/\sigma_\textnormal{tot}$ values are quite similar.
The discrepancies are not necessarily of concern, however, as A+16 select their sample based on stellar mass, while the sample presented here is based on the lens properties of the galaxies.

\bsp	
\label{lastpage}
\end{document}